\documentclass[aps,prd,twocolumn,nofootinbib]{revtex4}
\usepackage{amsmath,amssymb,graphicx}
\usepackage[usenames,dvipsnames]{color}
\usepackage{hyperref}
\newcommand{\be}{\begin{equation}}% can be used as {equation} or  {eqnarray}
\newcommand{\ee}{\end{equation}}
\newcommand{\beq}{\begin{eqnarray}}% can be used as {equation} or  {eqnarray}
\newcommand{\eeq}{\end{eqnarray}}
\begin{document}
%%%%%%%%%%%%%%%%%%%%%%%%%%%%%%%%%%%%%%%%%%%%%%%

\title{Unitarity in composite Higgs approaches with vector resonances}

\author{Daniele Barducci$^1$, Haiying Cai$^2$, Stefania De Curtis$^3$, Felipe J. Llanes-Estrada$^4$ and Stefano Moretti$^5$}
\affiliation{$^1$ LAPTH, Universit\'e de Savoie, CNRS, B.P.110, F-74941 Annecy-le-Vieux, France.\\
$^2$ Universit\'e de Lyon, F-69622 Lyon, France, Universit\'e Lyon 1, CNRS/IN2P3, UMR5822 IPNL,
F-69622 Villeurbanne Cedex, France.\\
$^3$ INFN, Sezione di Firenze, Via G. Sansone 1, 50019 Sesto Fiorentino, Italy.\\
$^4$ Departamento de F\'{\i}sica Te\'orica I, Univ. Complutense de Madrid,  28040 Madrid, Spain.\\
$^5$ School of Physics and Astronomy, University of Southampton, Southampton SO17 1BJ, U.K.}
\date{\today}

\begin{abstract}
\noindent
We examine a simple Composite Higgs Model (CHM) with vector resonances in addition to the Standard Model (SM) fields  in perturbation theory by using the $K$-matrix method to implement unitarity constraints. We find that the $W_LW_L$ scattering amplitude has an additional scalar pole (analogous to the $\sigma$ meson of QCD) as in generic strongly interacting extensions of the SM.
The mass and width of this dynamically generated scalar resonance are large
and the mass behaves contrary to the vector one, so that when
the vector resonance is lighter, the scalar one is heavier, and vice versa. We  also attempt an interpretation of this
new resonance. Altogether, the presence of the vector state with the symmetries of the CHM improve the low-energy unitarity behavior also in the scalar-isoscalar channel.

\begin{flushright}
\hspace{3cm} LAPTH-003/15\\
\end{flushright}
\end{abstract}

\maketitle

%%%%%%%%%%%%%%%%%%%%%%%%%%%%%%%%%%%%%%%%%%%%%%%%%
\section{Introduction}
%%%%%%%%%%%%%%%%%%%%%%%%%%%%%%%%%%%%%%%%%%%%%%%%%

The recent discovery of a Higgs-boson~\cite{ATLASCMS}
 has revived interest in the
Electro-Weak Symmetry Breaking (EWSB) sector of the Standard Model (SM) and beyond.
If this Higgs boson is confirmed to have exactly the couplings expected in the SM, a renormalizable theory of the EW interactions will be a closed chapter of physics history. Nevertheless, for several reasons, the particle physics community feels that there could be further new particles beyond the newly
discovered Higgs boson. It is then interesting that its reported mass, about 125 GeV, is of the same magnitude as the EW gauge bosons, $M_W\simeq 82$ GeV and $M_Z\simeq 91$ GeV, while no new particles have been seen up to 600--700 GeV. Particularly stringent are the bounds on possible further $W'$ or $Z'$ vector bosons and other particles coupling
to $WW$ and $WZ$ pairs below about 1.5 TeV~\cite{Khachatryan:2014gha}.

A natural scenario that theoretically fits this insight is that of a Composite Higgs Model (CHM) in which the Higgs state is a naturally light quasi-Nambu-Goldstone Boson (qNGB) stemming, like the longitudinal components of the gauge bosons $W_L$ and $Z_L$, from the spontaneous breaking of a higher energy symmetry~\cite{Kaplan:1983fs}.

While we do not really know what that symmetry might be like, Occam's razor dictates to examine first those models with the minimum number of ingredients. In the EWSB sector, this means the four Goldstone bosons that seem to be the low-energy content of the theory. A minimal such choice is the $SO(5)\to SO(4)$ breaking, proposed in~\cite{Agashe:2004rs}, that we spell out in section~\ref{sec:fields}.

Since our goal is to look forward to the TeV and multi-TeV region where new vector resonances may hide, and this is high-energy compared with the EW scale, we can profit from the Equivalence Theorem (ET)~\cite{ET} between the longitudinal $W_L$ components and the $\pi$ qNGB's.  The Lagrangian density that controls their low-energy interactions is discussed  in subsection~\ref{sec:effLag}.

 We then dedicate section~\ref{sec:pertscat} to the extraction of the scattering amplitudes among the low-energy particles in Leading Order (LO) chiral perturbation theory, extended by new vector resonances, that would correspond to the first accessible states (at the Large Hadron Collider (LHC)) of the CHM
considered here. The amplitudes therefore include contact chiral
interactions that are a polynomial in $s$ and Beyond the SM
(BSM) gauge-like interactions $\rho$-$\pi$-$\pi$
entering through $t$- and $u$-channel vector exchanges, with $\rho$ representing the accessible
(spin-1 gauge) resonances.
The polynomial terms imply strong interactions in spite of the
Higgs being light~\cite{Siringo:2001hm,Delgado:2013loa}. We find
that the $hh\to hh$ scattering amplitude vanishes in LO, thus we
calculate the other three (elastic $\pi^i\pi^j\to \pi^k\pi^l$ and
inelastic $\pi^i\pi^j\to hh$, $\pi^i\pi^j\to \pi^3 h$) relevant modes. Actually, we will prove that the latter scattering amplitude vanishes  due to a cancellation between the contributions from the two degenerate vector resonances contributing to the process.

The amplitudes are projected over the few lowest partial waves in section~\ref{sec:partialwv}, where we check the good convergence of the expansion at low-energy. While the vector channel is well behaved due to the new spin-1 resonances
introduced in the CHM scenario, this is not the case for the scalar-isoscalar partial wave:
we note the breakdown of unitarity by perturbation theory in the 2 TeV region for values of the parameters that are still compatible with current LHC bounds. It is well known, and continues being reinstated~\cite{Bhattacharyya:2012tj},
that, generically, if the couplings of the Higgs boson do not perfectly match the SM ones, unitarity violations are expected (see~\cite{Lahiri:2011ic} for an exception).

A traditional way out is to restrict the analysis to those values of the parameters $f$, $g_s$,
the `compositeness' (energy) scale and the new gauge coupling, respectively (that couple the new vector boson(s) to the longitudinal EW gauge sector), that allow perturbative unitarity to extend to relatively high scales \cite{Basso:2011na}, requiring for example partial Ultra-Violet (UV) completion so the couplings
cannot be arbitrarily strong~\cite{Contino:2011np}.

Instead, in this work, we focus section~\ref{sec:unitarity} on a non-perturbative model treatment of the partial waves by means of the $K$-matrix method,
irrespective of the value of the coupling. The unitarization methods start with rational instead of polynomial approximations to scattering functions and thus have no problem in incorporating strong resonant poles in the complex plane that limit the convergence of the polynomial ones.
They introduce some model dependence acceptable for exploratory analysis, which can however be reduced, at an increased level of sophistication, by basing them on dispersion relations or by directly working with the latter.

Since there are two relevant channels with distinct amplitudes, $\pi^i\pi^j\simeq W_LW_L, Z_LZ_L$ and $\pi^4\pi^4\equiv hh$, we employ a $2\times 2$ coupled- channel reaction matrix. Then in section~\ref{sec:rhorho} we take a preliminary look at the $\rho\rho$ threshold region, where the approach will require to be
extended to include a third channel. Since at those energies, above 2-3 TeV, the particle content of the theory typically becomes richer by a maze of new resonances, we refrain from performing an analysis beyond such energy point.

Our conclusions are  wrapped up in section~\ref{sec:fin}.

%%%%%%%%%%%%%%%%%%%%%%%%%%%%%%%%%%%%%%%%%%%%%%%%%
\section{Goldstone bosons and new vector fields
in a two-site model\label{sec:fields}}
%%%%%%%%%%%%%%%%%%%%%%%%%%%%%%%%%%%%%%%%%%%%%%%%%
\subsection{Particle and field content}
%%%%%%%%%%%%%%%%%%%%%%%%%%%%%%%%%%%%%%%%%%%%%%%
Let us, for simplicity, refer to the minimal  $SO(5)/SO(4)$ construct of a
CHM, which offers the minimum number of qNGBs filling the low-energy multiplet $(W_L^+,W_L^-,Z_L,h)\sim \vec{\pi}^T \equiv \left({\pi_1, \pi_2, \pi_3, \pi_4} \right)$,
and use it as a template to construct a typical effective Lagrangian
coupling vector resonances and qNGBs. The theoretical assumption is that  $\pi_4 =h $ can be identified as the light Higgs candidate and, as it becomes a qNGB, its couplings become predictable.

A convenient framework to implement spontaneous EWSB to LO in a chiral expansion is the non-linear sigma field formalism. The fifth ``$\sigma$''-like field that will acquire a high-energy Vacuum Expectation Value (VEV) breaking the symmetry is traded for a function of
$ \pi~=~\sqrt{\pi_1^2 + \pi_2^2+ \pi_3^2 +\pi_4^2}$.

The exponential representation $U =  {\exp}( i \sqrt 2 \pi^{\hat a } T^{\hat a} /f )$ naturally exposes the multiplet transformation under the global symmetry. Explicitly, in terms of the qNGBs,
\beq \label{unitrep}
 U = \left( \begin{array}{cc}
\textbf{1}_{4 \times 4} - \frac{
\vec \pi
\vec \pi^T}{\pi^2} \left(1 - \cos \frac{\pi}{f} \right) &  \frac{\vec \pi}{\pi} \sin \frac{\pi}{f} \\
 - \frac{\vec \pi^T}{\pi}\sin \frac{\pi}{f} & \cos \frac{\pi}{f}
\end{array} \right).  \eeq
Although we will not work with transverse gauge bosons in this contribution, soon setting their coupling to vanish,  $g_0=0$,
let us momentarily keep the covariant gauge derivative for completeness,
\beq
 D_\mu = \partial_\mu -i g_0 A_\mu^a  T^a.
\eeq

We will work with a so called ``two-site model'', where there are two sets of vector fields coupled to qNGBs.
The fields of the first site are the elementary gauge fields  $A_\mu$ with
\beq
A_\mu &=&  W_\mu^a T_L^a  + B_\mu  \delta^{ 3 a}  T_R^a \, ,
\eeq
where $T_L^a,\ T_R^a,\  a = 1,2,3 $ are the respective generators of $SU(2)_L$ and $SU(2)_R$. The $SU(2)\times SU(2)\simeq SO(4)$ symmetry remaining at this site is spontaneously broken to the custodial $SU(2)$  symmetry of the SM.

At the second, higher-energy site, there are additional vector fields
\beq\label{vector1}
\rho_\mu &=& \rho_{L,\mu}^a  T_L^a  + \rho_{R,\mu}^a  T_R^a   + a_{\mu}^{\hat a}  T^{\hat a} \ .
\eeq
These could be detected by means of the Drell-Yan or di-boson
processes~\cite{Agashe:2007ki,Agashe:2008jb,Barducci:2012kk}, since they are expected to couple to two fermions in an $s$-wave.  But, here, we will focus on their couplings to two vector bosons intervening in (longitudinal) gauge boson scattering.

The BSM generators of the coset space  $SO(5)/SO(4)$ are denoted with a hat over the adjoint index, $T^{\hat{a}},\ \hat a= 1, 2, 3, 4$, and $a_{\mu}^{\hat a} $ are the associated resonances while
the $\rho_{L/R,\mu}^a$ are those associated to $SU(2)_{L,R}$ (we will refer to them as vector resonances).
We may explicitly spell out the matrix representation of Eq.~(\ref{vector1}) as
\beq \label{vector2}
i  \rho_\mu^A T^A = \left( \begin{array}{c|c|c}
\frac{1}{2} \epsilon^{ijk} (\rho_{L, \mu}^k +\rho_{R, \mu}^k) & \frac{1}{2}  (\rho_{L,\mu}^i - \rho_{R,\mu}^i ) &  \frac{1}{\sqrt 2} {a_\mu^{\hat i}}   \\ \hline
 -  \frac{1}{2} (\rho_{L,\mu}^j - \rho_{R,\mu}^j )   & 0 &   \frac{1}{\sqrt 2} { a_\mu^{\hat 4} }   \\ \hline   -\frac{1}{\sqrt 2} {a_\mu^{\hat j} }  &  -\frac{1}{\sqrt 2} { a_\mu^{\hat 4}}  &0
\end{array} \right).\nonumber \\
\eeq

In the unitary gauge,  at low energy and after EWSB has occurred,
one can set $\pi_1 = \pi_2 =\pi_3 =0$, since they provide the $W_L$ and $Z_L$ components. At high energy it is more convenient to work with the
Goldstone fields and extract from them the $W_LW_L$ scattering amplitude via the
ET.

We thus employ all these pion-like fields and group them in a tensor parametrization
\beq \Pi  = \sqrt 2 {\pi ^{\hat a}}{T^{\hat a}} =  - i\left( \begin{array}{cc}
0_{4 \times 4}& \vec \pi \\
- {\vec \pi }^T &0
\end{array} \right)\, \label{pi}
\eeq
that is useful to construct couplings to the vector resonances.

The unitary representation in Eq.~(\ref{unitrep}) may be expressed as a product of two matrices of fields, one at each site, $ U = \Omega_1 \cdot \Omega_2 $.
These two matrices $\Omega_n,\ n=1,2$, are constructed from the $\Pi$ tensor in Eq.~(\ref{pi}) by the expressions
\beq
{\Omega _n} = 1 + i\frac{{{s_n}}}{\pi }\Pi  + \frac{{{c_n} - 1}}{{{\pi ^2}}}{\Pi ^2},\\ \nonumber
s_n = \sin (f\pi /f_n^2), ~~ {c_n} = \cos \left( {f\pi /f_n^2} \right),
\label{ome}
\eeq
where
\be \label{ff1f2}
f^2 =  f_1^2 f_2^2 / (f_1^2 + f_2^2)\ .
\ee
with $f_{1,2}$ being the ``pion decay constants'' associated to each of the two sites respectively.

The larger symmetry is spontaneously broken at the second site
$SO(5)_2\to SO(4)_2$ by a field $\phi_0^T = (0,0,0,0,1)$ that
acquires a VEV. The $SO(5)$ matrix  $\Omega_2$ can be used to arbitrarily orient
the direction of symmetry breaking, $\Phi_2 = \Omega_2 \phi_0$,
and this second-site field is then calculated using  Eqs.
(\ref{pi}) and (\ref{ome}) to yield
\beq \label{fieldphi2}
\Phi_2^T = \frac{1}{\pi} \sin(f \pi/f_2^2)  \left ( \pi_1 ,  \pi_2, \pi_3, \pi_4, \pi \cot(f \pi /f_2^2)  \right) .
\eeq
%

%%%%%%%%%%%%%%%%%%%%%%%%%%%%%%%%%%%%%%%%%%%%%%%%%%%%%%%
\subsection{Effective Lagrangian \label{sec:effLag}}
%%%%%%%%%%%%%%%%%%%%%%%%%%%%%%%%%%%%%%%%%%%%%%%%%%%%%%%
%
The coupling between the pion fields and the vector fields active at each of the two theory sites is affected by a minimum-coupling principle
introducing  covariant  derivatives,
\beq
D_\mu  \Omega_1 &=&  \partial_\mu \Omega_1   - i  g_0  A_\mu \Omega_1 +  i g_s  \Omega_1  \rho_\mu,   \nonumber  \\
D_\mu \Phi_2  &=&   \partial_\mu \Phi_2  - i g_s  \rho_\mu \Phi_2
\eeq 
where $g_s$ is the coupling strength associated with the new resonances.

Likewise, we introduce a field-strength tensor for the new
vector fields that allows the construction of a gauge-invariant
Lagrangian density (in the philosophy that there is a hidden gauge
symmetry)  as $\rho_{\mu \nu} = \partial_\mu  \rho_\nu -
\partial_\nu \rho_\mu  - i g_s  [ \rho_\mu,  \rho_\nu ] $.
 In our
application to ``low-energy'' $W_LW_L$ scattering we shall not
need the $\rho$ self-interactions (as a second such resonance
pushes the intermediate state containing it to yet higher
energies) and thus we will ignore the last term with the commutator,
keeping only the Maxwell kinetic energy term. An exception will be section~\ref{sec:rhorho} below, where we examine
the inelastic scattering $\pi \pi \to \rho \rho$ at a higher energy scale,
and the self-interaction $\rho$-$\rho$-$\rho$ needs to be included therein.

The resulting $\sigma$-model  Lagrangian is then
\beq\label{2siteL}
\mathcal{L}_{2 - site} &=&
 \frac{{f_1^2}}{4}Tr{\left({D_\mu }{\Omega _1}\right)^\dag {{D_\mu }{\Omega _1}}} + \frac{{f_2^2}}{2}{\left( {{D_\mu }{\Phi _2}} \right)^T}{D_\mu }{\Phi _2} \nonumber \\
&-& \frac{1}{4}Tr{\rho _{\mu \nu }}{\rho ^{\mu \nu }}.
\eeq

We should immediately acknowledge that the effective Lagrangian in Eq.~(\ref{2siteL}) does not contain the most possible general interactions. First, it is
built under the principle of a hidden gauge symmetry (using gauge theory as a template for the interaction of the new vector fields, when it is clear that new resonances may or may not be gauge bosons themselves). Second,
higher derivative, non-renormalizable counterterms should be added if further new physics lied at yet higher energy scales, though only relevant operators remain at low energy. And, additionally, we could mix the fields of the first site and the second site into a ``theory-space''
non-local term~\cite{DeCurtis:2011yx,DeCurtis:2014oza,Cai:2014kxa}  which is  allowed by the symmetries. This would be accomplished by defining $\Phi= \Omega_1 \Omega_2 \phi_0$ or
\beq
\Phi^T = \frac{1}{\pi} \sin(\pi/f)  \left ( \pi_1 ,  \pi_2, \pi_3, \pi_4, \pi \cot(\pi /f)  \right)
\eeq
in analogy with Eq.~(\ref{fieldphi2}),
that provides the additional two-derivative term
\beq \label{mixsites}
\mathcal{L}^{(2)}  &=&  \frac{f_0^2}{2} \left( {D^\mu }{\Phi } \right)^T {D_\mu }{\Phi }.
\eeq

In the rest of this article we will not pursue Eq.~(\ref{mixsites}) further but rather limit ourselves to the low-energy consequences of Eq.~(\ref{2siteL}) in the presence of relatively strongly coupled new vector fields. This allows us to neglect the transverse gauge bosons $W_T$, $Z_T$, turning off the  EW interaction,  i.e., $g_0 \to 0$.

We then consider $W_L W_L$ (through the ET, pion-pion) scattering. To reveal the $g_s$ content of Eq.~(\ref{2siteL})  we expand the two $\Omega$ fields, yielding
\beq
 \Omega_1 = \left( \begin{array}{cc}
1_{4 \times 4} - \frac{ f^2
}{ 2 f_1^4}  \vec \pi
\vec \pi^T   &  \frac{f }{f_1^2} {\vec \pi} (1-\frac{f^2}{6 f_1^4} \pi^2) \\
 -  \frac{f }{f_1^2} {\vec \pi^T}(1-\frac{f^2}{6 f_1^4} \pi^2)  &  1 - \frac{ f^2
}{ 2 f_1^4}   \pi^2
\end{array} \right) +  \mathcal{O} (\pi^4),
\nonumber \\
\eeq
\beq
 \Omega_2 = \left( \begin{array}{cc}
1_{4 \times 4} - \frac{ f^2
}{ 2 f_2^4}  \vec \pi
\vec \pi^T   &  \frac{f }{f_2^2} {\vec \pi}(1-\frac{f^2}{6 f_2^4} \pi^2)  \\
 -  \frac{f }{f_2^2} {\vec \pi^T} (1-\frac{f^2}{6 f_2^4} \pi^2) &  1 - \frac{ f^2
}{ 2 f_2^4}   \pi^2
\end{array} \right) +  \mathcal{O} (\pi^4)  \ .
\nonumber \\
\eeq

In the unitary gauge both the first and  second  term  in the Lagrangian density $\mathcal{L}_{2-site}$  of Eq.~(\ref{2siteL})  contribute to  the vertex $\rho\pi\pi$. In particular, the $L$ and $R$ vector couplings are unequal and separately listed. On the contrary, the coset resonances $\hat a_\mu$ defined in (\ref{vector1}), have only a small coupling to  $\pi^a$ induced after EWSB \cite{DeCurtis:2011yx}.

From the first term in $\mathcal{L}_{2-site}$ we obtain, with $i,j,k$ taking the values $1,2,3$,
\beq \label{firstpirho}
{\mathcal{L}_{\rho_L\pi\pi}}^{(1)} = \nonumber \\
\frac{{f^2 g_s}}{4 f_1^2}\left[ {{\varepsilon ^{ijk}}{\pi ^i}{\partial _\mu }{\pi ^j}\rho _{L\mu }^k + \left( {{\pi ^k}{\partial _\mu }{\pi ^4} - {\pi ^4}{\partial _\mu }{\pi ^k}} \right)\rho _{L\mu }^k} \right]  \nonumber, \\  \\
{\mathcal{L}_{\rho_R\pi\pi}}^{(1)}  =\nonumber \\
   \frac{{f^2 g_s}}{4 f_1^2}\left[ {{\varepsilon ^{ijk}}{\pi ^i}{\partial _\mu }{\pi ^j}\rho _{R\mu }^k - \left( {{\pi ^k}{\partial _\mu }{\pi ^4} - {\pi ^4}{\partial _\mu }{\pi ^k}} \right)\rho _{R\mu }^k} \right],  \nonumber \\
\eeq
and,  from the second term in $\mathcal{L}_{2-site}$,
\beq \label{secondpirho}
{\mathcal{L}_{\rho_L\pi\pi}}^{(2)}  =
\nonumber \\  \frac{{f^2 g_s}}{2 f_2^2}\left[ {{\varepsilon ^{ijk}}{\pi ^i}{\partial _\mu }{\pi ^j}\rho _{L\mu }^k + \left( {{\pi ^k}{\partial _\mu }{\pi ^4} - {\pi ^4}{\partial _\mu }{\pi ^k}} \right)\rho _{L\mu }^k} \right], \nonumber \\   \\
{\mathcal{L}_{\rho_R\pi\pi}}^{(2)}  = \nonumber \\
   \frac{{f^2 g_s}}{2 f_2^2}\left[ {{\varepsilon ^{ijk}}{\pi ^i}{\partial _\mu }{\pi ^j}\rho _{R\mu }^k - \left( {{\pi ^k}{\partial _\mu }{\pi ^4} - {\pi ^4}{\partial _\mu }{\pi ^k}} \right)\rho _{R\mu }^k} \right] . \nonumber \\
\eeq
While the 4$\pi$ vertices, with indices $a, b =1, 2, 3, 4$,  are collected as
\beq \label{4pivertex}
&& \frac{{f_1^2}}{4}Tr{\left( {{\partial _\mu }{\Omega _1}} \right)^\dag }{\partial _\mu }{\Omega _1} + \frac{f}{2}{\left( {{\partial _\mu }{\Phi _2}} \right)^\dag }{\partial _\mu }{\Phi _2}  \nonumber \\
 &\Rightarrow &
{\mathcal L}_{4\pi}=
\frac{f^4}{24f_1^6}\left[ {{{\left( {{\pi ^a}{\partial _\mu }{\pi ^a}} \right)}^2} - {{\left( {{\pi ^a}{\partial _\mu }{\pi ^b}} \right)}^2}} \right] \nonumber \\
 &+& \frac{f^4}{6f_2^6}\left[ {{{\left( {{\pi ^a}{\partial _\mu }{\pi ^a}} \right)}^2} - {{\left( {{\pi ^a}{\partial _\mu }{\pi ^b}} \right)}^2}} \right].
\eeq

The resulting effective interaction Lagrangian that combines Eq.~(\ref{firstpirho}) through (\ref{4pivertex}) can be employed in an energy range that is sufficiently above $2m_{W}\simeq 2m_h$ so that the ET applies and SM couplings are weaker than BSM couplings.
Since $\rho$ pair production is not described by the terms that we have kept in the $\pi\pi\to\pi\pi$ amplitudes, we also need to satisfy $\sqrt{s}<2m_\rho$, and we expect this scale to be similar to that intrinsic to the chiral expansion $4\pi f$, above which further derivative terms should also be included in
the $\pi\pi$ amplitude. Thus, the model Lagrangian can be of use in the energy range $(0.4,3)$ TeV for $m_\rho\simeq 2$ TeV. At the LHC, the low-energy end of this range is accessible and the polarization combination $W_L W_L$ can be activated.

The independent BSM parameters in the above Lagrangian density are three, namely $f_1$, $f_2$ and $g_s$, and they are related by the mass relations
\beq
&&m_{\rho_L}^2 = m_{\rho_R}^2 =g_s^2 f_1^2 /2 \label{KSFR},\\
&&m_{\hat a}^2= g_s^2 (f_1^2+f_2^2) /2,
\eeq
which hold true before the acquisition of a VEV by the Higgs field $h$ upon EWSB.
 Eq.~(\ref{KSFR})
is the so called KSFR relation~\cite{Kawarabayashi:1966kd}.
This  is not generally valid for BSM theories with an additional vector resonance but it is a consequence of the high symmetry imposed when the vector resonance is coupled to the Goldstone bosons as a gauge boson. Therefore, Eq.~(\ref{KSFR}) is a prediction of the theory in Eq.~(\ref{2siteL}). Notice also that the degenaration between $\rho_L$ and $\rho_R$ holds for $g_0=0$. Since we work in this approximation, we will take $m_{\rho_L} = m_{\rho_R}\equiv m_{\rho}$.

%%%%%%%%%%%%%%%%%%%%%%%%%%%%%%%%%%%%%%%%%%%%%%%%%%%%%%%%%%%%%%%%%%%%%
\section{Scattering amplitudes \label{sec:pertscat}}
%%%%%%%%%%%%%%%%%%%%%%%%%%%%%%%%%%%%%%%%%%%%%%%%%%%%%%%%%%%%%%%%%%%%%
\subsection{Tree-level amplitudes}
%%%%%%%%%%%%%%%%%%%%%%%%%%%%%%%%%%%%%%%%%%%%%%%%%%%%%%%%%%%%%%%%%%%%%
\begin{figure}
\includegraphics[width=0.45\textwidth]{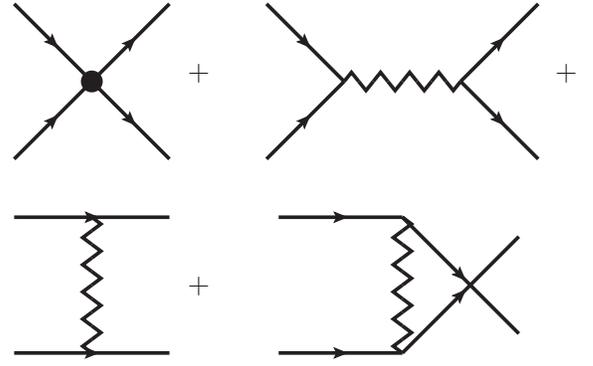}
\caption{\label{fig:Feynmandiags}Feynman diagrams that produce the tree-level amplitudes for pion ($W_L$) scattering in the energy range $2 m_{W,h}<<
\sqrt{s} << 4\pi f\sim$ several TeV.}
\end{figure}

We now extract the $\pi^i \pi^j \to \pi^k  \pi^l$ scattering amplitude
according to the Feynman diagrams in figure~\ref{fig:Feynmandiags}. This we denote as $A^{(\pi \pi\to \pi\pi )}$ or $A$ for short. Likewise we can obtain the inelastic $M^{(\pi \pi\to hh )}$,  $N^{(\pi \pi\to \pi h )}$, as well as the elastic $T^{(hh\to hh )}$ amplitudes, also shortened to $M$, $N$ and $T$, respectively. The elastic $hh$ scattering amplitude $T$ is not readily accessible at the LHC, since Higgs production rates are quite small and the final state reconstruction is quite poor, so that the final state $hh$ is rather unlikely and the initial state $hh$ does not abound either (owing to small Yukawa couplings to the proton content).
Additionally, in the considered CHM setup the $hh$ elastic amplitude vanishes in leading order in $s$, so that the BSM low-energy production must proceed from the $W_LW_L$ channel in the CHM.
(Nevertheless, the virtual re-scattering $W_LW_L\to hh\to W_LW_L$ can bring this dynamics into the visible $W_LW_L$ sector and some experimental information might be provided by the LHC, though a CLIC-type machine would be needed for a better study~\cite{Contino:2013gna}.)

Concentrating now on the elastic $\pi\pi$ amplitude, we note the following well-known isospin structure:
\beq
 A({\pi ^i}{\pi ^j} \to {\pi ^k}{\pi ^l}) = \nonumber \\
A(s,t,u) \delta ^{ij} \delta ^{kl} + A(t,s,u)\delta ^{ik} \delta ^{jl} + A(u,t,s)\delta ^{il} \delta ^{jk}\ .
\nonumber \\
\eeq

Standard calculation for the "two-site" model, considering the exchanges of $\rho_{L,R}$, with $m_{\rho_{L,R}}=m_\rho$ and neglecting terms {\cal O}($m_h^2/s$), leads to:
\beq \label{pipiamp}
A(s,t,u) =  \left( \frac{f^4 }{4 f_1^6} +\frac{f^4}{f_2^6} \right) s
\nonumber \\
-\frac{g_s^2}{2}  \left( \frac{f^2}{2 f_1^2}  + \frac{f^2}{f_2^2}  \right)^2 \left[\frac{s - u}{t - m_{\rho } ^2} + \frac{s - t}{u - m_{\rho } ^2}\right].
%\, .
\eeq
Notice that  $\rho_L$ and $\rho_R$ contribute to this amplitude with equal amounts, because the different sign in the Lagrangian densities of Eq.~(\ref{firstpirho}) and (\ref{secondpirho}) affects only vertices with $\pi^4=h$ that do not appear in tree-level $\pi^i\pi^j\to \rho_{L,R} \to \pi^k \pi^l$.

Since, in virtue of the co-called BRST identities \cite{BRST}, the interacting bosons are effectively spinless,
an efficient low-energy representation of the amplitude is obtained in terms of a few partial-wave projections.
We first project to definite isospin
\beq
A_0(s,t,u) &=&  3 A(s, t, u ) + A(t, s, u) + A(u,t, s),  \\
A_1(s,t,u) &=& A(t, s, u) - A(u, t, s), \\
A_2(s,t,u) &=& A(t, s, u)+ A(u, t, s)\, ,
\eeq
and then to definite $J=l$,
\beq \label{partials}
a^I_J(s) \,=\, \frac{1}{64\pi} \,\int_{-1}^{+1} {\rm d}\cos\theta\,
P_J(\cos\theta) \, A_I(s,t(s,\cos\theta))\, ,
\eeq
with $\cos\theta=1+ 2t/s$.

The resulting partial waves consist of two parts: a power expansion, coming from the purely $\pi\pi$ Lagrangian density, and terms due to $\rho$ exchange (at the lowest end of the $\sqrt{s}$ interval these could also be projected into low-energy derivative terms, but since we are exploring the model with  additional vector resonances, we keep the $\rho$ propagator explicitly).

Next, let us quote the inelastic scattering amplitude, that has weak isospin 0 by necessity, $M({\pi ^i}{\pi ^j} \to {h}{h}) = M{(s,t,u)}{\delta ^{ij}}$,
\beq \label{channelcoupling}
M(s,t,u) &=&
 \left( \frac{f^4 }{4 f_1^6} +\frac{f^4}{f_2^6} \right) (s - \frac{2}{3} m_h^2)  \\
&-& \frac{g_s^2}{2}  \left( \frac{f^2}{2 f_1^2}  + \frac{f^2}{f_2^2}  \right)^2 \left( \frac{s - u}{t - m_{\rho } ^2} + \frac{s - t}{u - m_{\rho } ^2} \right)
\nonumber \\
&-& ~ \frac{g_s^2}{2}  \left( \frac{f^2}{2 f_1^2}  + \frac{f^2}{f_2^2}  \right)^2  \frac{m_h^4}{m_\rho^2} \left( \frac{1}{t - m_{\rho } ^2} + \frac{1}{u - m_{\rho } ^2} \right)\! .   \nonumber
\eeq
Here the charged resonances $\rho_{L,R}^\pm$, which are degenerate with the neutral ones before EWSB and in the $g_0\to 0$ limit, are exchanged in the $t$ and $u$ channel.
This amplitude entails probability leak from the entry $W_LW_L$ channel to the (rather uncommon) $hh$ one. If, surprisingly, a large number of $hh$ events were visible at the LHC, this would point out to strong dynamics coupling this channel to $W_LW_L$. Notice that, for values of $m_\rho$ around 2 TeV or larger, $m_h$ and thus the last term in Eq.~(\ref{channelcoupling}) are negligible and $M(s,t,u)\simeq A(s,t,u)$.
Finally, let us consider the inelastic scattering amplitute $N( \pi^i \pi^j \to \pi^k h ) =N( s,t,u ) \delta^{ij}\delta^{k3}$,
\beq
N( s,t,u)  &=& \frac{g_s^2}{2} ( \frac{f^2}{2 f_1^2}
+ \frac{f^2}{f_2^2} )^2
\Big( \frac{u -t}{s - m_{\rho_L } ^2}\\ \nonumber && +\frac{s - u}{t - m_{\rho_L } ^2} + \frac{s - t}{u - m_{\rho_L } ^2}\Big)+(L\to R).
\eeq
Actually this amplitude
vanishes in this simple CHM. The reason for this is twofold. Firstly, the $\rho_L$ and $\rho_R$ exchange contributions to the amplitude are equal and with opposite sign, so they cancel due to the different sign in the Lagrangian densities of Eq.~(\ref{firstpirho}) and (\ref{secondpirho}) which affects vertices with $\pi^4=h$. Secondly, the chiral pion interactions for this process are known to start at order $p^6$ (we have only the low-energy $p^2$ terms plus those coming from vector boson exchange) and they require a violation of  discrete parities, as discussed in~\cite{Contino:2011np}. We thus ignore this channel in the following.

%%%%%%%%%%%%%%%%%%%%%%%%%%%%%%%%%%%%%%%%%%%%%%%%%%%%%%%%%%%%%%%%%%%%%
\subsection{Parameters}
%%%%%%%%%%%%%%%%%%%%%%%%%%%%%%%%%%%%%%%%%%%%%%%%%%%%%%%%%%%%%%%%%%%%%

To shorten notation in Eqs.~(\ref{pipiamp}) and (\ref{channelcoupling}), and in view of the standard factor in the partial wave projection in Eq.~(\ref{partials}), it is useful to employ two constants $K_1$ and $K_2$ defined as
\beq \label{defKs}
K_1= \frac{1}{16\pi} \left(\frac{f^4}{4 f_1^6}+\frac{f^4}{f_2^6}\right) ,\nonumber\\
K_2= \frac{g_s^2}{16\pi} \left( \frac{f^2}{2 f_1^2}  +
\frac{f^2}{f_2^2}  \right)^2
\eeq
(notice that $K_1$ is dimensionful).

For the minimal CHM to have only the $\rho_{L,R}$   and no further resonances in the low-energy region, we need to split up the coset resonances: this is achieved by  requiring $f_2\to \infty$. This simplifies Eqs.~(\ref{defKs}), since from Eq.~(\ref{ff1f2}) $f\to f_1$, to read
\beq \label{Ks_sc1}
K_1= \frac{1}{32\pi} \frac{1}{2 f^2} ,\\  
K_2= \frac{g_s^2}{64\pi} =\frac{1}{32\pi}\frac{m^2_\rho}{f^2}\ .
\eeq

In this limit, the model has only two parameters,
$f$ and $g_s$ (or equivalently $f$ and $m_\rho$ after using Eq.~(\ref{KSFR})), that can be immediately obtained once the amplitudes (functions of $K_1$ and $K_2$) become known, by solving $g_s=8\sqrt{\pi K_2}$ and
$f=\frac{1}{8\sqrt{\pi K_1}}$.
Nevertheless, there is a third parameter, necessary to regulate the pole of the vector resonance, $\Gamma_\rho$, that is generally independent of the other two, $f$ and $m_\rho$.

Because of Eq.~(\ref{width}) below, $K_2$ could be traded for the common partial $\pi \pi$ widths of the new vector particles $\Gamma_{\rho\pi\pi}$
(in fact $\rho_{L,R}$ have the same mass in the degenerate limit and the same couplings to $\pi \pi$ as it is clear from Eqs.~(\ref{firstpirho}-\ref{secondpirho})). Note that $\Gamma_\rho\simeq\Gamma_{\rho\pi\pi}$ if there were no extra strongly coupled fermions  to which the vector particles could have a sizeable decay amplitude, for example, because they would be as heavy as or heavier than the $\rho$ itself. But, if instead there were such fermions, what would appear in the propagators is the total width $\Gamma_\rho$ as opposed to the partial $\pi\pi$ width in Eq.~(\ref{width}) \cite{Barducci:2012kk}. It is also true that  in CHMs, the top quark has a composite component (via the partial compositeness mechanism) and, as a consequence, it is sizeably coupled to the new vector resonances leading to a 
non-negligible $t \bar t$ decay channel.
Thus we have $\Gamma_\rho$ as a free parameter of our analysis 
but, for moderate values of the order a few percent,
it  does not make a large difference except in the vector-isovector channel itself (of course, in other channels the vector state is exchanged in $t$ and $u$ diagrams, where its width just slightly modifies low-energy potentials).\color{black}

In our plots, both $f$ and $m_\rho$ will be taken as relatively small,
for which the effects we are describing are relevant at the LHC run II but are not yet discarded by LHC run I studies. Particularly on $f$ there are somewhat more stringent bounds from LEP, but these are based on loop-computations that are to be taken with a grain of salt because  they (currently)  cut-off the virtual effects of any new high-energy physics. 

We contrast this three-parameter scenario with a more complete one, e.g., the 4-Dimensional CHM (4DCHM) of
Ref.~\cite{DeCurtis:2011yx},
with a finite $f_2$ corresponding to non-decoupled coset resonances but sufficiently heavy to be undetected by the LHC.
Even if the coset resonances are only smoothly coupled to $\pi\pi$ so that we can neglect their contribution in the tree level amplitudes, their presence is indirectly manifest by the dependence of $K_1$ and $K_2$ on $f_2$.
In this second scenario there are up to four free parameters, that can be taken as $f$,  $g_s$, $m_\rho$ and $\Gamma_\rho$.
%%%%%%%%%%%%%%%%

With one or the other scenario we should have enough flexibility to describe in an effective way many models that just have vector resonances at low energy in addition to the already known EWSB  sector including a SM-like Higgs state. For example,  the Higgs-like boson under study would be a dilaton (so that there are further strong interactions in the EWSB sector) and there is an additional vector boson. Another possibility would be that
the B-L seemingly accidental symmetry was actually a gauge symmetry, which would bring about an additional vector boson but without the relation in Eq.~(\ref{ff1f2}).

For convenience, the parameter content for the two scenarios is summarized in Table~\ref{table:params}.
\begin{table}
\caption{
Dependent and independent parameters in the two scenarios considered in our numeric computations.  $K_1$, $K_2$, $m_\rho$ and $g_s$ directly appear in our amplitudes and could thus be reconstructed from experiment in principle. \label{table:params}
}
\begin{tabular}{|ccc|}\hline
Parameter & Scenario 1      & Scenario 2       \\
         & ($\hat a_\mu$ decoupled) & (complete 4DCHM) \\ \hline  \hline
$f_1$     &   $=f$          & =$\sqrt{2} m_\rho/g_s$                \\
$f_2$     &   $\infty$     & $\frac 1 {f_2^2}=\frac 1 {f^2} -\frac 1 {f_1^2}$               \\ \hline
$f$       & indep. variable & indep. variable               \\
$m_\rho$ & indep. variable & indep. variable  \\
$g_s$    & $=\sqrt{2}m_\rho/f$ & indep. variable           \\
$\Gamma_\rho$& indep. variable & indep. variable  \\ \hline
$K_1$    & $=\frac{1}{16\pi}\frac{1}{4f^2}$ & $ K_1(f,m_\rho,g_s)$  \\
$K_2$    & $=\frac{1}{16\pi}\frac{m_\rho^2}{2f^2}$ & $K_2(f,m_\rho,g_s)$ \\
\hline
\end{tabular}

\end{table}

%%%%%%%%%%%%%%%%%%%%%%%%%%%%%%%%%%%%%%%%%%%%%%%%%%%%%%%%%%%%%%%%
\section{Elastic partial waves \label{sec:partialwv}}
%%%%%%%%%%%%%%%%%%%%%%%%%%%%%%%%%%%%%%%%%%%%%%%%%%%%%%%%%%%%%%%%

We now quote the lowest non-vanishing partial waves for each isospin channel from Eq.~(\ref{partials}). These are
\beq \label{isoscalarwave}
a^0_{0}(s) =  \nonumber \\
K_1 s
+  K_2  \bigg[\, \left(\frac{m_{\rho}^2}{s}+2 \right) \log \left(1+ \frac{s}{m_{\rho }^2}\right)-1 \, \bigg]\, , \nonumber \\
\eeq
\beq \label{isovectorwave}
a^1_{1}(s) = \nonumber \\
\frac{K_1}{6} s +  \frac{K_2}{6 s^2 (m_\rho^2-s)}
\bigg[\ -s(6m_\rho^4+6m_\rho^2s-13s^2)
 \nonumber \\
+3( 2m_\rho^6+3m_\rho^4 s -3m_\rho^2 s^2 -2s^3)
\log\left(1+\frac{s}{m_\rho^2} \right) \bigg] \nonumber \\
\eeq
and
\beq \label{isotensorwave}
a^2_{0}(s) = \nonumber \\
\frac{-K_1}{2} s -  \frac{K_2}{2}
\bigg[\, \left(\frac{m_{\rho}^2}{s}+2 \right)
\log \left(1+\frac{s}{m_{\rho }^2}\right)  -1 \, \bigg]  \nonumber \\
\eeq
(the first and third ones satisfy $a_0^2(s) = -\frac{1}{2}a_0^0(s)$).
These are the lowest-order partial wave projections in each isospin channel:  higher ones are suppressed by an additional power of $s$. For example, in the isoscalar channel, the tensor ($l=2$) partial wave starts as
\beq
a^0_{2}(s) =
\frac{K_2 }{s^2 } \left(\frac{m_\rho^2}{s}+2\right)\bigg[\ -3s(2m_\rho^2+s)
\nonumber \\
+(6m_\rho^4+ 6m_\rho^2s + s^2)\log\left(1+\frac{s}{m_\rho^2} \right) \bigg], \nonumber \\
\eeq
i.e., without the linear term $\propto K_1 s$.

Let us delve in the amplitudes for a few lines.
First, we notice that the isotensor partial wave $a^2_0$ is repulsive at low energy, as Eq.~(\ref{isotensorwave}) has an explicit negative sign, while the
other two channels are attractive.
Thus, if a doubly charged resonance couples $W^+W^+$, this means that this partial wave probably changes sign (to avoid violating Wigner's causality bound), driven by higher order chiral terms, and the convergence of the series will be very poor.

Our second observation is that, naturally, the vector-isovector partial wave in Eq.~(\ref{isovectorwave}) presents a simple pole at $s=m_\rho^2$.
Of course, this singularity is just a feature of perturbation theory blindly applied:  if we resum the $\pi\pi$ bubble insertions in the $\rho$ propagator, the BSM-vector width $\Gamma_\rho$ naturally regulates the denominator. Then, one should substitute $\frac{1}{m_\rho^2-s}$ by
$\frac{1}{(m_\rho-i\Gamma_\rho/2)^2-s}$.
This isovector partial wave is the only one that acquires an imaginary part at this stage; the others remain real and are thus in violation of unitarity, that is only satisfied in perturbation theory by proceeding to the next order. 
Again, in the case in which the fermion decay channels are suppressed,
the width is dominated by the  partial width $\Gamma_{\rho\pi\pi}$, otherwise it is an independent parameter.

A straightforward calculation yields
\be \label{width}
\Gamma_{\rho\pi\pi} = m_\rho \frac{K_2}{6},
\ee
equation that provides a beautiful interpretation of the chiral constant $K_2$ in terms of $\Gamma_{\rho\pi\pi}/m_\rho$ which simply becomes
\be
\Gamma_{\rho\pi\pi} =\frac {m_\rho^3}{192 \pi f^2}
\ee
in scenario 1, where one can eliminate $g_s$.

An important observation
is that the contribution of the BSM vector resonance to Eq.~(\ref{isoscalarwave}) is positive. For $s<m_\rho^2$, in the low-energy regime, the factor
\beq  \label{logexpansion}
\bigg[\
\left(\frac{m_{\rho}^2}{s}+2 \right) \log \left(1+ \frac{s}{m_{\rho }^2}\right)-1 \, \bigg]\ \sim \frac{3s}{2m_\rho^2} > 0.
\eeq
Thus, at low energy, Eq.~(\ref{isoscalarwave}) becomes
\beq \label{a00limit}
a^0_{0}(s) &=&
\left(K_1 + \frac{3K_2}{2 m_\rho^2}
\right) s.
\eeq

The ratio of the two terms in this expression happens to be, for $m_{\hat{a}}$ not too far from $m_\rho$ (else we are in scenario 1),
$3(m_{\hat{a}}^2-m_\rho^2)/(2m_\rho^2)$.
So, if $m_{\hat{a}} >  m_\rho$ and the two
states are not closely degenerate (which would invalidate our treatment
anyway because an explicit $\hat{a}$ resonance would have to enter the
amplitudes), both terms contribute to the low-energy theorem.

By using the explicit expressions for $K_1$ and $K_2$ in Eq.~(\ref{defKs}) plus   $m_\rho$ in Eq.~(\ref{KSFR}), we get the low-energy $(s<m_\rho^2)$ behaviour for  the scalar partial wave,
\beq\label{low}
a^0_{0}(s) &\simeq& \frac{s}{16\pi f^2}\frac{f^6}{f_1^6} \left[ \big(\frac{1}{4} + \frac{f_1^6}{f_2^6}\big)+\frac{3}{4}\big(1+2 \frac{f_1^2}{f_2^6}\big)^2\right]\\ \nonumber
&\simeq& \frac{s}{16\pi f^2},
\eeq
with the first contribution from the four-pion contact terms and the second from the $\rho$-exchange terms. These two contributions sum up to the expected low-energy result for $a_0^0$ being regulated only by the symmetry breaking scale $f$. It is easy to check it in scenario 1 ($f_2\to\infty,~f_1=f$),  but it holds true for any choice of $f_2$ which satisfies the relation in Eq.~(\ref{ff1f2}).
This result does not depend on the $\rho$ mass or coupling, so at this order  the scale of unitarity saturation is  totally controlled by $f$.
Actually, at the next order there is some amelioration (the unitarity scale is pushed higher) because the logarithm in Eq.~(\ref{logexpansion}) is an alternating series and the next term is negative, slightly reducing $f$. This happens at order $s^2/m_\rho^4$,
\be\label{a00NLO}
a_0^0 \simeq \frac s {16 \pi f^2}  - s^2 \left(
\frac{2}{3} \frac{K_2}{m_\rho^4} \right).
\ee

%%%%%%%%%%%%%%%%%%%%%%%%%%%%%%%%%%%%%%%%%%%%%%%%%%%%%%%%%%%%%%
\subsection{Matching to low-energy effective theory}
%%%%%%%%%%%%%%%%%%%%%%%%%%%%%%%%%%%%%%%%%%%%%%%%%%%%%%%%%%%%%%

From the LHC run I data other investigators have extracted some bounds on the low-energy coefficients of operators extending the SM in the language of effective theory, that is being used profusely, especially in the non-linear representation of the sigma model~\cite{Alonso:2012px,Pich:2013fba,Degrande:2012wf,Buchalla:2013rka}.

 We can profit from this approach to reduce the parameter space that needs to be explored. Taking again the low-energy limit of our expression for $a_0^0$ in Eq.~(\ref{a00NLO}),
we can identify the leading term in the commonly used expression~\cite{Delgado:2013hxa}
\be
a_0^0 \simeq \frac{1}{16\pi v^2}(1-a^2) s
\ee
where currently $a\in (0.7,1.3)$ is not excluded by the LHC run I~\cite{Baak:2012kk} (in the weakly coupled SM, $a=1$ and this strong amplitude vanishes).
Since the $s$ coefficient is positive, the relevant bound for us is the lower one, $a\geq0.7$.

By comparing with Eq.~(\ref{a00NLO}), in our CHM
$a^2=1-v^2/f^2$. For large $m_\rho$ so that $s^2/m_\rho^4$ may be neglected, we have
\be \label{boundK1}
f \geq 350~{\rm GeV}\ .
\ee
(If $m_\rho$ is kept finite  and the Next-to-LO
(NLO) term in Eq.~(\ref{a00NLO}) is not negligible, then the bound is a little bit  less stringent: for example, for $s=m^2_\rho/2$ the bound is lowered to $\sim 320$ GeV).
Since one naturally expects $f>v$
we will explore this range of values of $f$ in the following. We do not employ precision EW constraints here since they are contingent on what new physics enters through loop corrections \cite{Baak:2012kk}.

Next we compare the inelastic amplitude $m_0^0$ for $\pi\pi\arrowvert_{I=0}\to hh$ between the actual model and the low-energy effective theory. For this, let us quickly reorder the amplitude in Eq.~(\ref{channelcoupling})  to expose it as a power series in $m_h$, from which we will keep only the zeroth order term since we are also neglecting $M_W$ and $M_Z$ that are of the same order in any sensible counting,
\beq \label{channelcoupling2}
M(s,t,u) &=&
 16\pi K_1 s  - 8\pi K_2 \left( \frac{s - u}{t - m_{\rho } ^2} + \frac{s - t}{u - m_{\rho } ^2} \right)
\nonumber \\
&-& \frac{32\pi}{3} K_1 m_h^2
\nonumber \\
&-& ~ 8\pi K_2  \frac{m_h^4}{m_\rho^2} \left( \frac{1}{t - m_{\rho } ^2} + \frac{1}{u - m_{\rho } ^2} \right)\! .   \nonumber
\eeq
In the massless limit we immediately note the equality with the elastic amplitude $M(m_h=0) = A(s,t,u)$.
To obtain the isospin-zero projection we note that the Clebsch-Gordan coefficients rotating
$\arrowvert \pi^a\pi^b \rangle$
to $\arrowvert \pi\pi\rangle_{I=0}$ bring in factors of $1/\sqrt{3}$ so that
\begin{eqnarray}
M_{0} &=& \frac{1}{\sqrt{3}} \sum_a M(\pi^a\pi^a \to hh) \nonumber \\
         &=& \sqrt{3} M.
\end{eqnarray}
Then the scalar partial wave projection of that cross-channel amplitude $M_0$ becomes
\be \label{inelasticprojection}
m_0^0 = \frac{1}{64\pi} \int_{-1}^1 dx M_0(s,t(s,x)).
\ee
Performing the integral we find $m_0^0\simeq \frac{\sqrt{3}}{2} a_0^0$ in the limit $m_h\simeq 0$ (that is, much smaller than $s$ and $m_\rho$).
The proportionality factor is easy to understand at least for small $s$. Just note that to project the elastic $A$ over zero isospin we used $A_0 = 3A(s) + A(t) +A(u)$ and that, since $3s+t+u\simeq 2s$, $A_0 = 2 M$ for small $s$.
Finally the $\sqrt{3}$ comes from $\sqrt{3}^{-1} \sum_a \delta^{aa}$ and reflects the different final state in $A$ ($\pi^a\pi^b$) and $M$ ($hh$).

Comparing now with the non-linear version of the Higgs Effective Field Theory (EFT), where
\be
m_0^0= \frac{\sqrt{3}}{32\pi v^2} (a^2-b) s + O(s^2)\ ,
\ee
it follows that
\be
\left( K_1 +\frac{3K_2}{m_\rho^2} \right) = \frac{a^2-b}{16\pi v^2}
\ee
and, by using eq.~(\ref{low}), that
\be
\frac{1}{16\pi f^2} = \frac{a^2-b}{16\pi v^2}\ .
\ee

It further follows  that $a$ and $b$ (independent parameters of the EFT) are correlated in CHMs by
\be
(a^2-b) = (1-a^2)
\ee
that relates the strength of elastic scattering beyond the SM (RHS) with the inelastic one (LHS).

At higher order ${\cal O}(s^2)$ in the expansion, strong vector resonances appearing in $W_LW_L$ scattering leave  sizeable $a_4$ and $a_5$ coefficients. We do not pursue the topic further here but refer to~\cite{Espriu:2014jya}
where the low-energy parameter map is studied in detail with attention to the appearance or not of a BSM vector resonance.

%%%%%%%%%%%%%%%%%%%%%%%%%%%%%%%%%%%%%%%%%%%%%%%%%%%%%%%%%%%%%%
\subsection{Numerical results}
%%%%%%%%%%%%%%%%%%%%%%%%%%%%%%%%%%%%%%%%%%%%%%%%%%%%%%%%%%%%%%
\begin{figure}
\includegraphics*[width=0.45\textwidth]{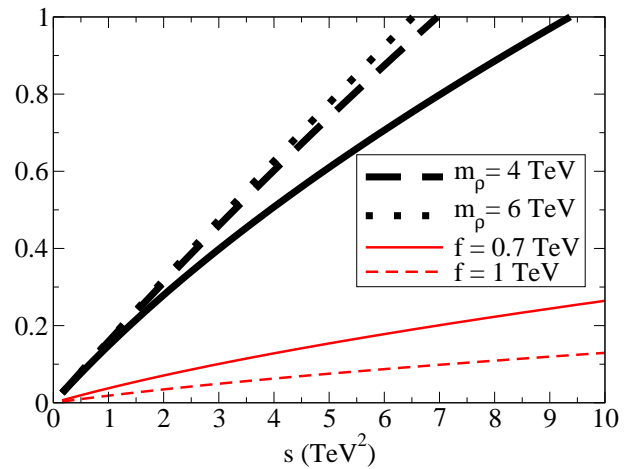}
\caption{\label{fig:a00pert}  Modulus of the
scalar-isoscalar partial wave $a^0_0$ (a real, positive number for real $m_\rho$ and $s$) in Eq.~(\ref{isoscalarwave}) as function of Mandelstam-$s$.
Solid line: reference values of the parameters, $m_\rho=2$ TeV, $f=350$ GeV, with the thickness of the line representing the uncertainty in the width $\Gamma_\rho\in(5-20\%)$. Thick dashed and dotted lines: same with $m_\rho=4$ and 6 TeV.
Red thin lines: instead, increase $f$ to $0.7$ and 1 TeV (at the larger $\Gamma_\rho=0.2m_\rho$ width value and $m_\rho=2$ TeV ).}
\end{figure}

We now numerically examine the elastic $\pi^i\pi^j\to \pi^k\pi^l$ amplitudes as function of the three independent parameters $f$, $m_\rho$ and $\Gamma_\rho$ of scenario 1 ($\hat{a}_\mu$-decoupled).

In figure~\ref{fig:a00pert} we show the $a_0^0$ scalar-isoscalar scattering amplitude as function of Mandelstam-$s$.
We choose as reference parameter set $m_\rho=2$ TeV, $f=350$ GeV and $\Gamma_\rho=20\% m_\rho$ (thick, solid line).
To show the dependence on parameters, the thick dashed and dotted lines correspond to increasing $m_\rho$ to 4 and 6 TeV respectively. The red thin ones towards the bottom of the plot correspond to $m_\rho=2$ TeV and increasing $f$ to 0.7 and 1 TeV instead.
The width of the vector state exchanged in the $t$, $u$ channels is of little concern for this scalar-channel amplitude. The thickness of the line itself corresponds to varying the width between 5 and 20\%.
The unitarity bound $|a_0^0| \leq 1$ is violated at around 3 TeV, invalidating perturbation theory. This happens at lower energies for larger $m_\rho$ and at higher energies for larger $f$: for the higher $f$ values shown, the violation of unitarity happens at higher scales between 4 and 5 TeV.

\begin{figure}
\includegraphics*[width=0.45\textwidth]{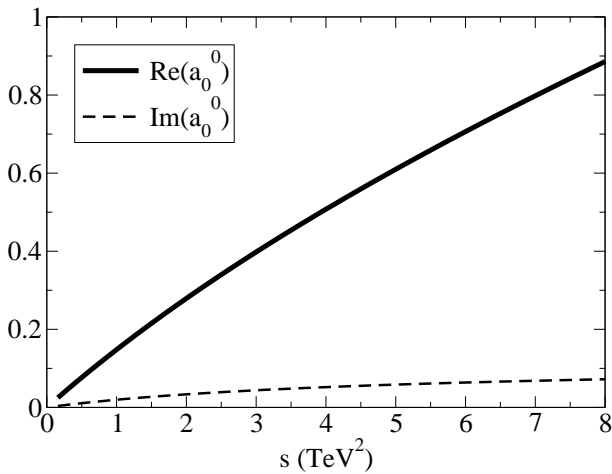}
\caption{\label{fig:complexa00}
We plot the scalar-isoscalar partial wave in Eq.~(\ref{isoscalarwave})
explicitly showing the real and imaginary parts. The later comes from
the substitution $m_\rho\to (m_\rho -i \frac{\Gamma_\rho}{2})$ with $m_\rho=2$TeV, $\Gamma_\rho=20\%m_\rho$, $f=350$ GeV.}
\end{figure}

One effect in the perturbative amplitude of adding the width of the vector resonance (induced by its potentially large coupling to fermions, for example) is the appearance of an imaginary part in the $a_0^0$ amplitude because of the substitution $m_\rho\to (m_\rho -i \frac{\Gamma_\rho}{2})$ in Eq.~(\ref{isoscalarwave}).  In figure~\ref{fig:complexa00} we separately plot  the real and imaginary parts of $a_0^0$.
Though now complex, $a_0^0$  still fails the unitarity test, that would only be satisfied perturbatively if an NLO amplitude was added: the induced imaginary part is too small.

Since adding or not a width $\Gamma_\rho$ will not change any of our qualitative statements, particularly in figure~\ref{fig:polemovement1} below, we will subsequently fix $\Gamma_\rho$ in Eq.~(\ref{isoscalarwave}) to the larger value $\Gamma_\rho/m_\rho=20\%$ when dealing with the scalar channel in this subsection, and neglect $\Gamma_\rho$ altogether afterwards (in the vector channel it does make a difference as we explain next).

The isotensor wave $a^2_0$ is repulsive and thus not expected to resonate at low energy, and since its value is $-a_0^0/2$ we do not plot it explicitly.

\begin{figure}
\includegraphics*[width=0.45\textwidth]{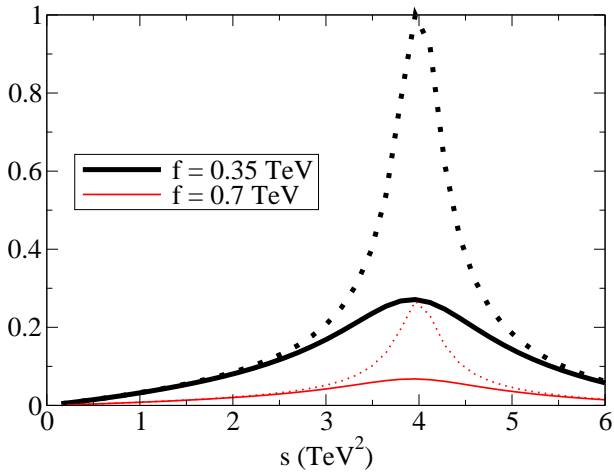}
\caption{\label{fig:a11pert} Modulus of the vector-isovector partial wave $|a_1^1|$ in Eq.~(\ref{isovectorwave}) as function of Mandelstam-$s$. Solid lines: $\Gamma_\rho=20\% m_\rho$. Dotted lines: $\Gamma_\rho=5\% m_\rho$ for $m_\rho=2$ TeV.}
\end{figure}

The vector-isovector wave is shown in turn, again in perturbation theory, in figure
\ref{fig:a11pert}.
Unitarity is here perfectly respected in all the low and intermediate energy region up to a few TeV, and it is saturated of course at $s=4\ {\rm TeV}^2$ (since we have fixed $m_\rho=2$ TeV to exemplify), where $\arrowvert a_1^1\arrowvert =1$ for the narrower resonance. Other values of $m_\rho$ trivially displace the pole; but it is worth showing how this channel reacts to $\Gamma_\rho$, to which we assign the two values 5\% and 20\%. We also display the calculation for two values of $f$. It is seen that the peak is more prominent for small $f$, while larger values of $f$ tend to make it disappear (this is because of the smaller $K_2$, coupling intensity of the resonance to the $\pi\pi$ channel, at fixed $m_\rho$).

\begin{figure}[h]
\includegraphics*[width=0.45\textwidth]{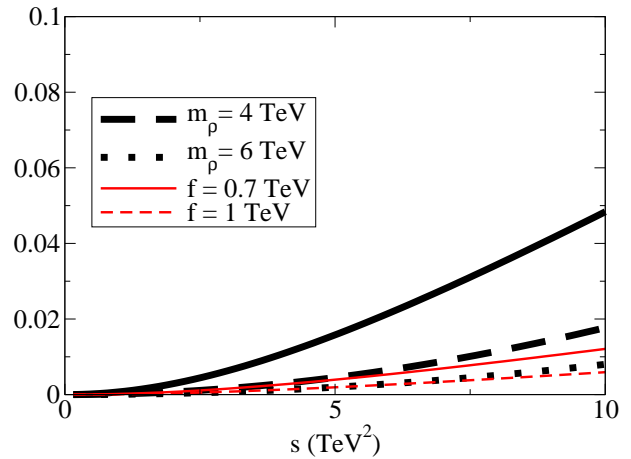}
\caption{\label{fig:a02pert} Tensor-isoscalar partial wave $a_2^0$. Lines as in figure~\ref{fig:a00pert}.}
\end{figure}

Figure~\ref{fig:a02pert} displays in turn the tensor-isoscalar amplitude $a_2^0$.
This partial wave is seen to be very small (note the scale in the $y$-axis has been divided by 10), far from reaching the unitarity bound $|a_2^0|=1$, and thus perturbative and non-resonating. In comparing with the $J=0$ wave in figure~\ref{fig:a00pert}, we see that the convergence of the partial wave expansion of the amplitude in the low-energy region is excellent, with the isoscalar channel practically dominated by the lowest, scalar partial wave. This is analogous to the QCD situation where the scalar $\sigma$ pole at 450 MeV dominates low-energy $\pi\pi$ scattering, with the first tensor resonance, the $f_2(1270)$, being much higher in mass.
A difference between the scalar and tensor channels is their reaction to increasing $m_\rho$. While in the scalar channel this makes the amplitude
larger bringing unitarity violation to a smaller scale, in the tensor channel it makes it smaller just like increasing $f$ does.

\begin{figure}[h]
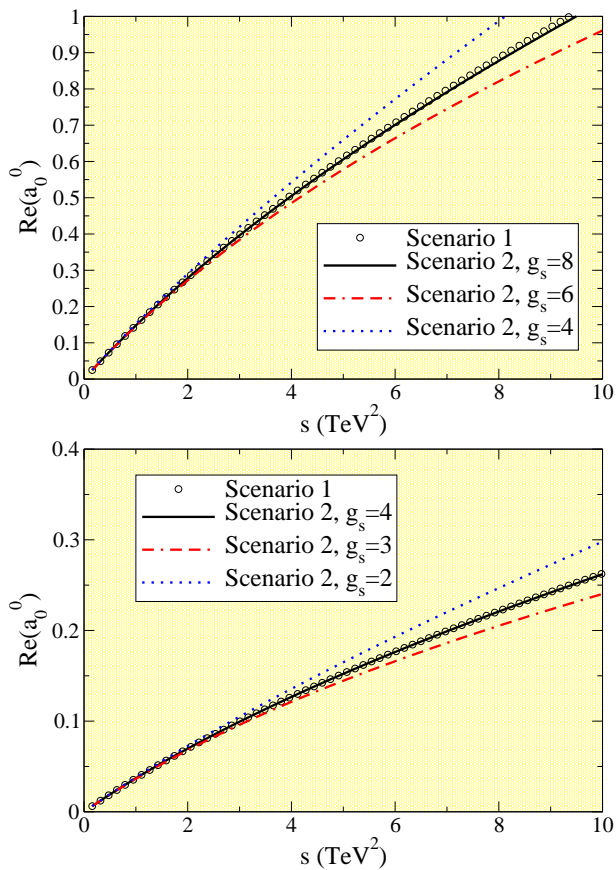

\includegraphics*[width=0.45\textwidth]{FIGS.DIR/CompareScenarios12_f0.35.eps}
\includegraphics*[width=0.45\textwidth]{FIGS.DIR/CompareScenarios12_f0.7.eps}
\caption{\label{fig:comparescpert}
Scenario 2 converges towards scenario 1 for a certain $g_s$. With the parameters here ($m_\rho=2 $ TeV, $\Gamma_\rho=20\%m_\rho$), this happens for $g_s$ slightly larger than 8 if $f=0.35$ TeV (top plot) and for $g_s$ a bit above 4 for $f=0.7$ TeV (bottom).  
}
\end{figure}

We now turn to the second scenario from table~\ref{table:params}.
We shade the plots with numerical data from this scenario 2 (soft yellow online) to easily distinguish them.
In figure~\ref{fig:comparescpert} we compare both scenarios. We have fixed $f$, $m_\rho$ and $\Gamma_\rho$ and vary only $g_s$ in scenario 2, which is a free parameter controlling the coupling of the vector to the $\pi\pi$ channel (the remaining width presumably due to fermion couplings).
As can be seen, the results at low energy are not too disparate, and the second scenario converges towards the first when $m_{\hat{a}}
\to \infty$ (the largest such mass in the plot is about 16 TeV and corresponds to the $g_s=4$ curve).

\begin{figure}[h]
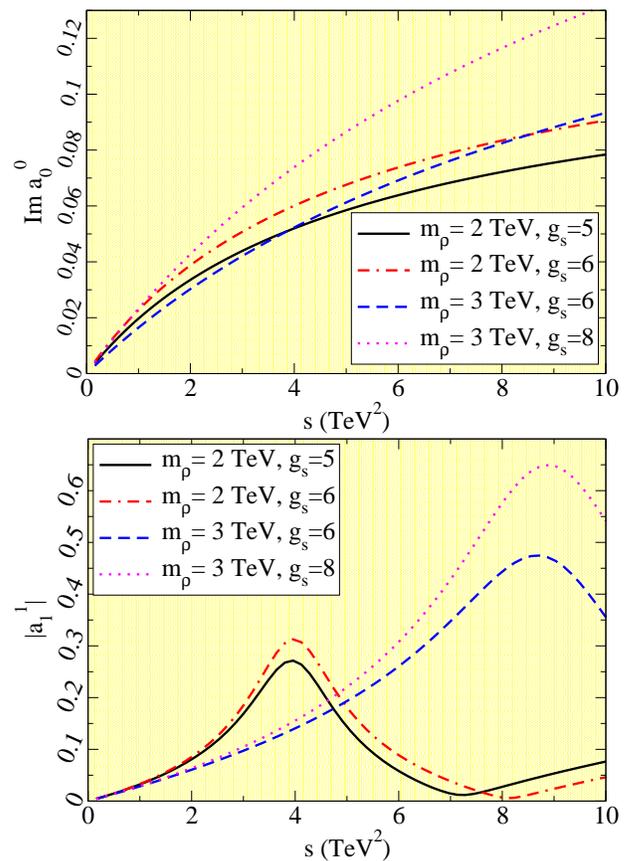

\includegraphics*[width=0.45\textwidth]{FIGS.DIR/Perta00_sc2.eps}
\includegraphics*[width=0.45\textwidth]{FIGS.DIR/Perta11_sc2.eps}
\caption{\label{fig:sc2pert}
Scenario 2, the complete 4DCHM. Top: imaginary part of the scalar, isoscalar partial wave. Bottom: modulus of the vector, isovector one. Here, $f=350$ GeV  and $\Gamma_\rho/m_\rho=20\%$.
}
\end{figure}

In figures~\ref{fig:comparescpert} and~\ref{fig:sc2pert} we stay with scenario 2, assuming  the 4DCHM without decoupling the axial vector resonances.
The first of them offers a comparison with scenario 1 in the limit $m_{\hat{a}} \to \infty$, achieved for a finite value of $g_s$ {for which $f_2\to\infty$.}
As seen in the plot, the convergence is good though not monotonic in sign. In any case, the two scenarios seem to give comparable results for both values of $f=0.35$ and 0.7 TeV.

Further detail  is provided by figure~\ref{fig:sc2pert}.
Here $f=350$ GeV, so the imaginary part of the scalar channel in the top plot is directly comparable with the real part in the top plot in figure~\ref{fig:comparescpert}. In the real part, unitarity is clearly violated by exceeding the bound $|a_0^0|\leq 1$.
The imaginary part shows additionally that the relation $Im a_0^0 = |a_0^0|^2$
is not satisfied even when the bound is not exceeded  (the imaginary part is of order 0.01-0.02 in the low-energy region).

For the shown values of $g_s$, $m_{\hat{a}}$ is either half a TeV or a TeV higher than $m_\rho$, so its effect in the low-energy physics is less prominent, but it is not totally decoupled as in scenario 1 {(we have not included the  $\hat{a}$ exchanges in our computation because, as stated in section II, the $\hat a$ have small couplings to $\pi \pi$ induced by EWSB and their contribution to the unitarity dynamics is negligible).}

The lower plot in figure~\ref{fig:sc2pert} shows the vector-isovector wave that behaves unsurprisingly, peaked at the nominal mass, with the width that we have fixed a priori, and with a strength that grows with $g_s$, its coupling to the $\pi\pi$ channel (when $\Gamma_{\pi\pi}\to \Gamma_\rho$ the peak height of the modulus approaches 1, saturating unitarity).

\color{black}
%%%%%%%%%%%%%%%%%%%%%%%%%%%%%%%%%%%%%%%%%%%%%%%%%%%%%%%%%%%%%%
\section{Coupled channel unitarity \label{sec:unitarity}}
%%%%%%%%%%%%%%%%%%%%%%%%%%%%%%%%%%%%%%%%%%%%%%%%%%%%%%%%%%%%%%
\subsection{Analysis for physical $s$}
%%%%%%%%%%%%%%%%%%%%%%%%%%%%%%%%%%%%%%%%%%%%%%%%%%%%%%%%%%%%%%
An unpleasant feature of perturbation theory is the breakdown of unitarity that can be catastrophic if the interactions become relatively strong,
even surpassing the unitarity bound.
This limits the reach of effective low-energy Lagrangians, but dispersion-relation based analysis provides a way around. There are several tools and methods of varying sophistication to address unitarity, but for this exploration
we adopt the simplest, so called ``K-matrix'' method~\cite{Wigner:1946zz}. In its original form, this guarantees unitarity but not the appearance of a proper right cut, so we use a slightly modified version, sometimes called ``improved K-matrix'' approach. It is based on the observation that the often appearing loop function
\be \label{bubble}
J(s) = \frac{-1}{\pi}\log\left( \frac{-s}{\Lambda^2} \right)
\ee
provides a right-hand cut in the complex-$s$ plane for $s\in(0,\infty)$. Here $\Lambda$ is an appropriate high-energy cutoff that we can naturally take as $\Lambda=m_\rho$ to analyze the lower-energy scalar channel.

If the amplitude $M(\pi^i\pi^j\to\pi^4\pi^4)$ vanished, we could unitarize the elastic $A(\pi^i\pi^j\to \pi^i\pi^j)$ scalar amplitude as
$\tilde a = a (1-Ja)^{-1}$. This amplitude would satisfy ${\rm Im}\ a =|a|^2$, but mixing with the Higgs-Higgs channel
introduces the inelastic scalar $m^0_0$ projection of Eq.~(\ref{inelasticprojection}) in this relation.
This happens only in  the isospin-zero channel where there is mixing between the
$W_L W_L$ and Higgs-Higgs states because of the non-vanishing channel-coupling amplitude in Eq.~(\ref{channelcoupling}).
Thus, we expect a probability leak from the
$\pi^i\pi^j$ to the $\pi^4\pi^4$ channels.

Under this circumstance, the exact elastic unitarity relation that the amplitude needs to satisfy is
\be \label{unitcheck}
{\rm Im}\ a =|a|^2 +|m|^2
\ee
(the $0$ indices are omitted). A convenient way to implement it is to construct a reaction matrix that contains both channels in perturbation theory,
\be
k=\left(
\begin{tabular}{cc}
$a$ & $m$ \\ $m$ & 0
\end{tabular} \right)
\ee
(noticing the vanishing of the Higgs-Higgs elastic amplitude in LO perturbation theory, a model feature).
This perturbative $2\times 2$ reaction matrix can be unitarized by
\be \label{unitarizedk}
\tilde k = k (1 -J k)^{-1}
\ee
if $k$ is small, which happens at low $s$, so this model amplitude reproduces the LO perturbative behavior since $\tilde k\simeq k+\dots$ therein.

The unitarization prescription of the $K$-matrix is by no means unique, with alternatives being the large-$N$ treatment, the Inverse Amplitude Method (IAM), or the $N/D$ ansatz~\cite{Oller:1999mg}, but all yield qualitatively similar results in the scalar channel over the right cut (the physical $s$ region) and nearby in the complex plane.

The matrix element $\tilde k_{11}\equiv \tilde a$ thus substitutes $a_0^0(s)$ for all but the lowest energies. Its explicit expression is
\be
\tilde a =\frac{a + J m^2}{1-Ja-J^2m^2} \ .
\ee

Eq.~(\ref{unitcheck}) is now satisfied exactly as long as the perturbative $a$ is real. Since $\Gamma_\rho$ was shown to induce a small imaginary part in the perturbative scalar amplitude, there is a residual unitarity problem of that same order. To avoid it, and since the effect of the width was numerically small in the scalar channel, we neglect $\Gamma_\rho$ here altogether.

\begin{figure}[h] %%%%%%%%
\includegraphics*[width=0.45\textwidth]{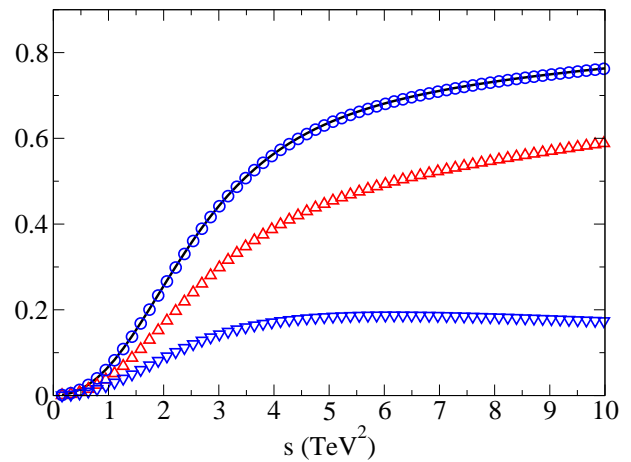}
\caption{\label{fig:unitaritytest}
Test of unitarity for
${\rm Im}\ \tilde a = {\rm Im}\ \tilde k_{11}$ (black solid line).
Bottom line (downward triangles, blue online):
$|\tilde m|^2= |\tilde k_{12}|^2$.
Second from bottom (upward triangles, red online):
 $|\tilde a|^2= |\tilde k_{11}|^2$. Circles (on top of the imaginary part): the sum of the last two, unitarity compliant. Here $\Gamma_\rho/m_\rho=5\%$, for larger values a small difference is visible due to the perturbative $a_0^0$ acquiring an imaginary part.
}
\end{figure}

In figure~\ref{fig:unitaritytest} we present one of the three possible checks of unitarity for the $K$ matrix (in scenario 1 for definiteness), showing the satisfaction of Eq.~(\ref{unitcheck})
for the  $\tilde a=\tilde k_{11}$ and $\tilde m=\tilde k_{12}$ quantities (the other two independent checks are also satisfied, but not shown).
     %%%%%%%%

By comparing the lowest two curves in the figure one can see that after unitarization the loss of probability from the $W_LW_L$ channel to the $hh$ channel is still about 25\% of the elastic scattering one.

\begin{figure}[h]
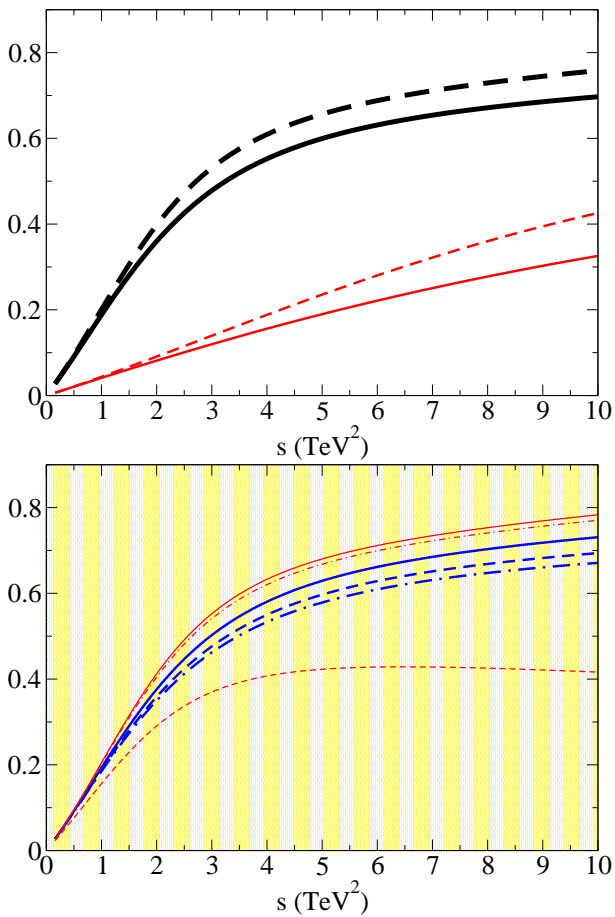

\includegraphics*[width=0.45\textwidth]{FIGS.DIR/k00_sc1.eps}
\includegraphics*[width=0.45\textwidth]{FIGS.DIR/k00_sc2.eps}
\caption{\label{fig:unitarizedk}
Modulus of the unitarized elastic matrix element $|\tilde k_{11}|=|\tilde a_0^0(s)|$, that remains below 1 in all the energy interval of interest.
First plot (scenario 1): solid lines correspond to $m_\rho=2$ TeV, dashed ones to $m_\rho=4$ TeV. From thicker to thinner, $f=0.35,0.7$ TeV respectively.
Lower plot: in scenario 2, we fix $f=$0.35 TeV and $\Gamma_\rho/m_\rho=5\%$.
The thick lines (blue online) correspond to $m_\rho=2$ TeV, with $g_s=4$ (solid), 6 (dash-dotted) and 8 (dashed). The thin ones (red online) were in turn calculated with $g_s=6$ (solid), 8 (dash-dotted) and 10 (dashed), and all have $m_\rho=4$ TeV.
}
\end{figure}

In figure~\ref{fig:unitarizedk} we then present the modulus of the elastic amplitude (after unitarization) $|\tilde k_{11}|=|\tilde a_0^0(s)|$
that shows how the goal has been met: the amplitude equals the LO perturbation theory for the lowest $s$ but later moderates its growth satisfying the theoretical constraints.
In both scenarios it is apparent that the presence of a $\rho$-boson at low energy and strongly coupled to the $\pi\pi$ channel weakens the strength of the scalar channel and makes it more perturbative.

Other unitarization methods will lead to qualitatively similar predictions. To make an appreciable gain in accuracy, if ever necessary, the complete NLO amplitudes would have to be calculated and then fed into the more sophisticated IAM~\cite{GomezNicola:2001as}
 (that requires both LO and NLO).

\begin{figure}[h]
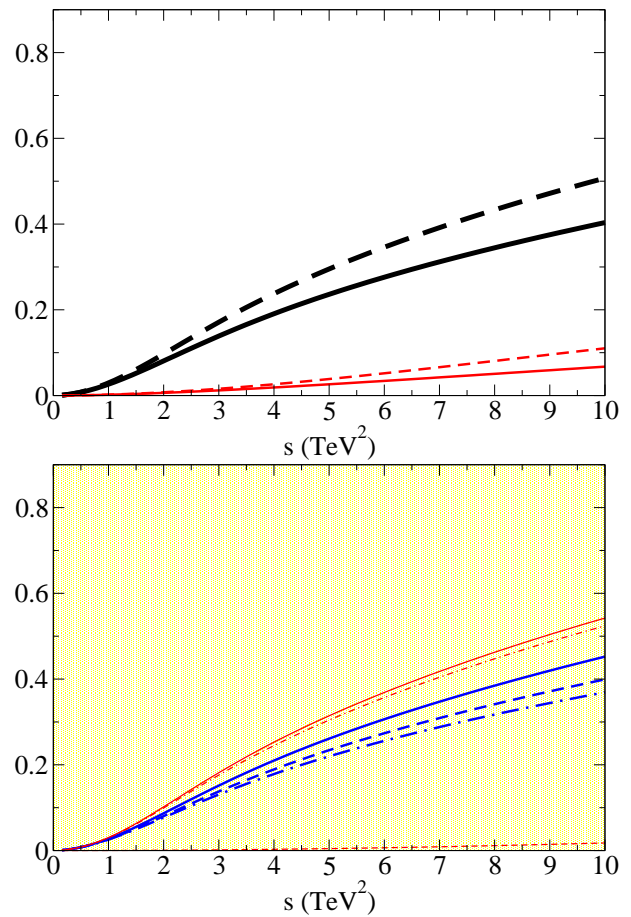

\includegraphics*[width=0.45\textwidth]{FIGS.DIR/k22_sc1.eps}
\includegraphics*[width=0.45\textwidth]{FIGS.DIR/k22_sc2.eps}
\caption{\label{fig:unitarizedk22}
Same as figure~\ref{fig:unitarizedk} but for $|\tilde k_{22}|=|\tilde t_0^0(s)|$ the elastic $hh\to hh$ scattering that vanishes in the perturbative amplitude at LO. Unitarization requires this channel to have finite probability too.
}
\end{figure}

An interesting feature that illustrates the limitations of perturbation theory is presented in figure~\ref{fig:unitarizedk22}, that shows the modulus $|\tilde k_{22}|=|\tilde t_0^0(s)|$ of the elastic $hh\to hh$ or $\pi^4\pi^4\to \pi^4\pi^4$ scattering amplitude for zero angular momentum.
It is remarkable that the scattering amplitude takes a finite and indeed non-negligible value when it is zero in LO perturbation theory. This reflects in the figure in that the linear term near the origin is zero, but the amplitude quickly overcomes this and takes appreciable values. This effect occurs, of course, by re-scattering through the other channel, $hh\to W_LW_L\to hh$, and since the unitarization procedure typically resums the imaginary part of all such re-scatterings, it is able to yield a finite value even when only LO perturbation theory is at hand.

%%%%%%%%%%%%%%%%%%%%%%%%%%%%%%%%%%%%%%%%%%%%%%%%%%%%%%%%%%%%%%
\subsection{Extension to the complex $s$-plane: $\sigma$ pole}
%%%%%%%%%%%%%%%%%%%%%%%%%%%%%%%%%%%%%%%%%%%%%%%%%%%%%%%%%%%%%%

The most important mid-range attraction of the nucleon-nucleon potential in nuclear physics is controlled by an exchange with scalar quantum numbers, that is usually assigned to a ``$\sigma$'' particle. In the modern understanding  of QCD, this particle, perhaps too broad to be called as such, is a resonance or
pole in the second Riemann sheet of the scattering amplitude $\pi\pi\to \pi\pi$ with a mass of about $450$ MeV. It is now known with remarkable precision thanks to the use of accurate dispersion relations with a wealth of low and mid-energy data.
The strong interaction that we observe in $W_LW_L$ scattering in figure~\ref{fig:unitarizedk}  also comes from an equivalent pole in the second Riemann sheet.
\begin{figure}[h]
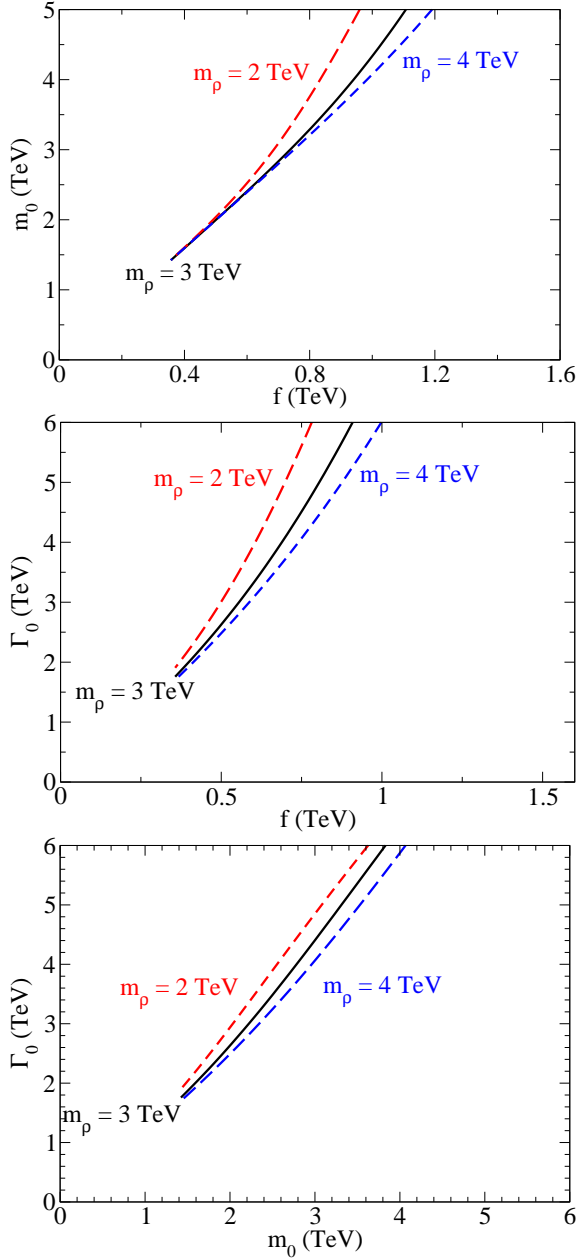

\includegraphics*[width=0.42\textwidth]{FIGS.DIR/Massofsigmavs_f_sc1.eps}
\includegraphics*[width=0.42\textwidth]{FIGS.DIR/Widthofsigmavs_f_sc1.eps}
\includegraphics*[width=0.42\textwidth]{FIGS.DIR/MassofSigmavsWidthofSigma_sc1.eps}
\caption{\label{fig:polemovement1}
Top and middle plots: evolution of the mass and width of the dynamically generated scalar $\sigma$-like resonance as function of $f$ in the limit $f_2\to\infty$ (scenario 1), for three values of the vector mass. Bottom: width of the scalar resonance against its mass, with $f$ being now just the parameter of the trajectory in this plane.
}
\end{figure}
To expose it with the $K$-matrix method (obviously a model, thus less precise than the Roy equations~\cite{Ananthanarayan:2000ht,Kaminski:2006qe}
 that can later be applied when/if data becomes available) we extend the variable $s$ to the complex plane in our computer code. The extension to the second Riemann sheet, where resonance poles in the lower-half plane can appear (since they are forbidden in the first sheet due to causality) is implemented in the
loop function in Eq.~(\ref{bubble}).
It is sufficient to take the logarithm to be cut in $(-\infty,0)$ (so the argument is defined between $-\pi$ and $\pi$) and exploit the simple prescription
\be \label{secondsheet}
\log \left(\frac{-s}{m_\rho^2}\right)_{\rm II} =
\log \left({\rm Abs}\left(\frac{s}{m_\rho^2}\right)\right)+
i \left({\rm Arg}\left(\frac{s}{m_\rho^2}\right)- \pi\right).
\ee

We then employ the CERN standard minimization program MINUIT to search the complex plane for zeroes of the determinant of $(1-Jk)$ that yield the poles of
the unitarized scattering amplitude $\tilde k$, in accord with Eq.~(\ref{unitarizedk}). We find exactly one such pole and interpret its position as an `effective' resonance with a certain mass and width, given by $\sqrt{s}= M_0 -i \Gamma_0/2$ (where the $0$ reminds us of its apparent spin).

\begin{figure}[h]
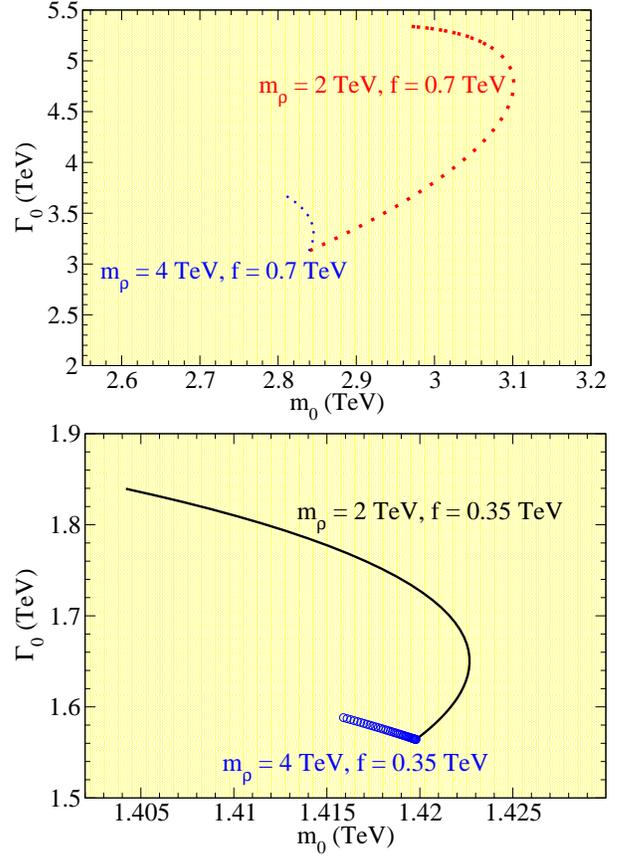

\includegraphics*[width=0.44\textwidth]{FIGS.DIR/MassofSigmavsWidthofSigmaf0.7_sc2.eps}
\includegraphics*[width=0.44\textwidth]{FIGS.DIR/MassofSigmavsWidthofSigmaf0.35_sc2.eps}
\caption{\label{fig:polemovement2}
Varying $g_s$ produces a curve in the $(M_0,\Gamma_0)$ plane
of the $\sigma$-like pole as extracted from its complex-$s$ position, for $f=0.7$ TeV (top) and $f=0.35$ TeV (bottom), and for $m_\rho=2$ or 4 TeV as indicated.  It may happen that for each value of $M_0$ there are two values of $\Gamma_0$. This reflects the oscillation with the parameter of the curve, $g_s$, visible in figure~\ref{fig:polemovement3}, and caused by the non-trivial dependence of Eq.~(\ref{defKs}) on $g_s$.
}
\end{figure}

\begin{figure}[h]
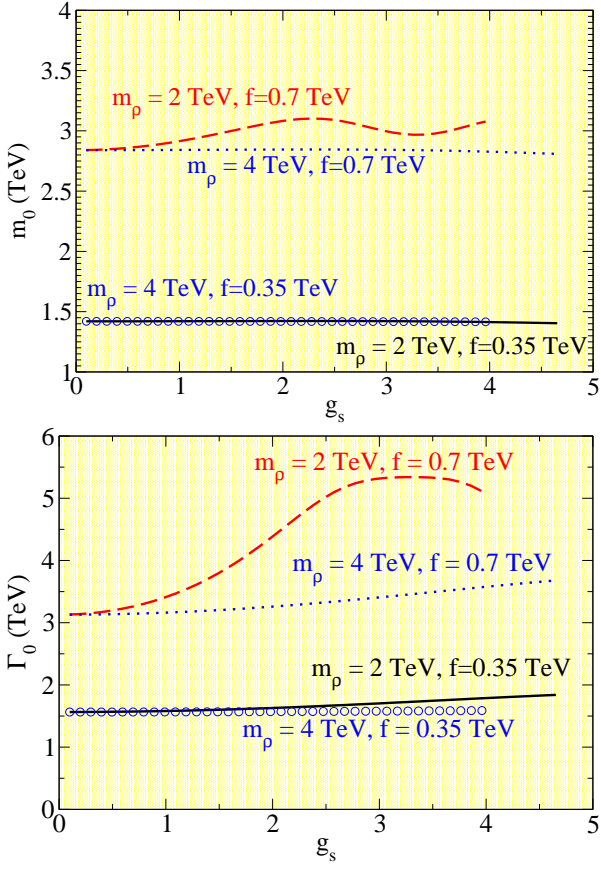

\includegraphics*[width=0.44\textwidth]{FIGS.DIR/Massofsigmavs_gs_sc2.eps}
\includegraphics*[width=0.44\textwidth]{FIGS.DIR/Widthofsigmavs_gs_sc2.eps}
\caption{\label{fig:polemovement3}
Evolution of the mass (top) and width (bottom) of the dynamically generated scalar $\sigma$-like resonance as function of $g_s$ in scenario 2, for
fixed vector masses 2 and 4 TeV and fixed $f=0.35$ and 0.7 TeV as indicated.
}
\end{figure}

 We have tracked the evolution of this pole in scenario 1 for fixed
$\Gamma_\rho/m_\rho=0.05$ as a function of $f$ and for three values of
$m_\rho$ as shown in figure~\ref{fig:polemovement1}.
We show the behavior of the mass and width of the scalar pole in terms of $f$ in the first two plots, then eliminate this parameter to see directly $\Gamma_0(m_0)$ in the third plot.

As the LHC data advances and perhaps tightens the constraint on $f$ rising its minimal allowed value, the possible positions of the pole recede in the complex plane (larger $M_0$ and $\Gamma_0$ for larger $f$, at fixed $m_\rho$).

If, in contrast, the vector resonance relatively decouples from the low energy Goldstone bosons,  because of its heavy mass (large $m_\rho$), we recover the known broad pole from generic strongly interacting theories a bit under 2 TeV. As the $\rho$ becomes lighter, this pole moves up in energy (for fixed $f$) and becomes broader.
Since its width is similar in size to its mass (bottom plot) for all values of $f$ (bottom plot of figure~\ref{fig:polemovement1}) its interpretation as an unstable particle is as difficult as in QCD.

We now turn to  scenario 2 (the full 4DCHM with a finite $m_{\hat{a}}$ mass, though we still neglect its exchange) and plot the result in figure~\ref{fig:polemovement2} and~\ref{fig:polemovement3}.

The top plot of figure~\ref{fig:polemovement2} shows the complex-plane evolution (with $g_s$ being the curve's parameter) for fixed $f=0.35$ TeV and $m_\rho=2$ or 4 TeV, while the bottom one corresponds to the same $m_\rho$ but a larger $f=0.7$~TeV.
In comparing with the bottom plot of figure~\ref{fig:polemovement1} we see that the dependence of the mass and width on $f$ and $m_\rho$ is similar:
larger $f$ entails a heavier and {broader} pole, while
a heavier  $m_\rho$ moves it the opposite way, towards lighter values of $m_0$ and $\Gamma_0$.
The pole position bends gently in the complex plane, which must translate into a slight oscillation as function of $g_s$. This is observed in figure~\ref{fig:polemovement3} where we represent $M_0$ and $\Gamma_0$ as function of $g_s$.

The most noticeable effect is that the mass is practically independent of $g_s$ except for larger $f$ and smaller $m_\rho$, where it fluctuates somewhat more. The curves in both figures end when $g_s$ is so large that either $f_2$  {is very large} or $\Gamma_{\rho\pi\pi}$ equals its maximum possible value
$\Gamma_\rho$ that we have set in both cases to 5\% of $m_\rho$.

The width of the scalar pole moves slightly up with that of the vector pole, but in all the effect of $g_s$ is not striking (for moderate widths of the vector resonance itself $\Gamma_\rho$).

So the conclusion from both scenarios is how the $\sigma$ pole recedes deeper in the complex plane as $f$ is increased, and its mass  behaves opposite to the vector one for fixed couplings.

Thus, if the vector resonance is
coupled as the symmetries of the CHM dictate, it not only unitarizes the vector channel with its quantum numbers, but it also improves perturbative unitarity in the scalar channel 
(not generically true for other low-energy vector resonances).
%%%%%%%%%%%%%%%%%%%%%%%%%%%%%%%%%%%%%%%%%%%%%%%%%%%%%%%%%%%%%%
\section{$\rho\rho$ production}\label{sec:rhorho}
%%%%%%%%%%%%%%%%%%%%%%%%%%%%%%%%%%%%%%%%%%%%%%%%%%%%%%%%%%%%%%

We now  make a slight extension of our low-energy study and
take a look at  the $\rho\rho$ threshold. Double-$\rho$ intermediate states should  be taken into account in the elastic $\pi\pi$ amplitude, but we will not recalculate those and leave it to future investigation. Nevertheless, for completeness,  we find interesting to explore  the inelastic scattering amplitude $B(\pi^a \pi^b \to \rho_L^c  \rho_L^d)$, containing $h$,$\pi$ and $\rho$ exchange channels, and with a general isospin structure:
\beq
B^{ab\to cd} = \nonumber \\
A(s,t,u){\delta ^{ab}}{\delta ^{cd}} + B(s,t,u){\delta ^{ac}}{\delta ^{bd}} + B(s,u,t){\delta ^{ad}}{\delta ^{bc}}. \nonumber \\
\eeq
We quote explicit expressions for $A(s,t,u)$ and $B(s,t,u)$, where
threshold effects will be important (if the threshold can be reached at all), and are encoded in the  phase-space velocity factor $\beta_\rho = \sqrt{1-4m_\rho^2/s}$~\cite{Cai:2013ira}.

\beq
A(s,t,u) =   \nonumber \\
\frac{g_s^2}{4 } \left( \frac{f^2}{2 f_1^2}  + \frac{f^2}{f_2^2}  \right)^2 \frac{1}{ m_\rho^2 \beta_\rho^2 } \frac{1}{u } \left(  \frac{s}{2}\left(\beta_\rho^2 +1\right)+t
 -m_\rho^2\right)^2 \nonumber \\
 +   \frac{g_s^2}{4 } \left( \frac{f^2}{2 f_1^2}  + \frac{f^2}{f_2^2}  \right)^2 \frac{1}{ m_\rho^2\beta_\rho^2} \frac{1}{t} \left(  \frac{s}{2}\left(\beta_\rho^2 -1\right) -t
 + m_\rho^2\right)^2 ~,\nonumber  \\   \\
B(s,t,u) = \nonumber \\
\frac{g_s^2}{4 }\!\! \left( \frac{f^2}{2 f_1^2}  + \frac{f^2}{f_2^2}  \right)\!\! \left[\frac{(s+2 m_\rho^2)(t - u)}{(s - m_{\rho} ^2) m_\rho^2} +\left( \frac{f^2}{2 f_1^2}  + \frac{f^2}{f_2^2}  \right) \frac{1}{m_\rho^2\beta_\rho^2} \right. \nonumber \\ \cdot\!\!   \left.
\left( \frac{\left(  \frac{s}{2}\left(\beta_\rho^2 +1\right)+t -m_\rho^2\right)^2}{u}
 +  \frac{\left(  \frac{s}{2}\left(\beta_\rho^2 -1\right) -t + m_\rho^2\right)^2}{(t-m_h^2)} \right)\!\! \right]~. \nonumber \\
\eeq
As  we have  $m_h \ll m_\rho$, it is justified to take the massless $m_h\to 0$ limit in this amplitude.
\begin{figure}[h]
\includegraphics*[width=0.45\textwidth]{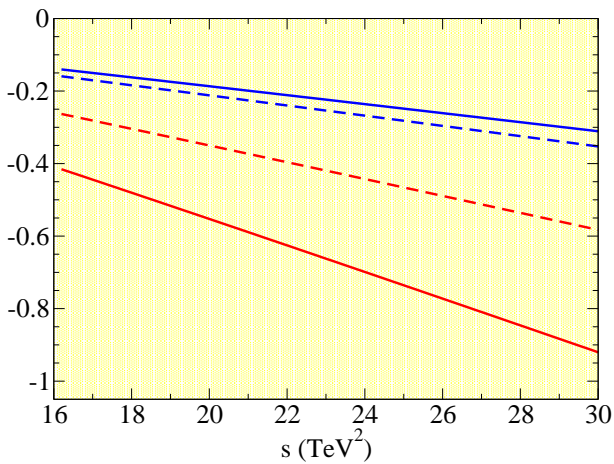}
\caption{\label{fig:b00} Scalar-isoscalar perturbative amplitude of 
$\pi\pi\to \rho_L\rho_L$ for  $m_\rho=2$ TeV, with  $g_s =3 $ (dashed line) and  $g_s=4$ (solid line) in scenario 2. The two bottom lines and the two top lines correspond respectively to $f=0.35$ and 0.7 TeV.}
\end{figure}
Since the LHC might reach the $\rho\rho$ threshold but would probably be energy and luminosity constrained to go much above, in practice, we need consider only the scalar partial wave (higher ones being suppressed by powers of $p^l$).
After projecting, and setting $m_h=0$, we get\\
\beq \label{rhoprod}
b^0_0(s)= \nonumber \\
K_2
\frac{5}{8\beta_\rho^2 m_\rho^2} \bigg[  \left( 1-2 \beta_\rho ^2 \right) s -2 m_\rho^2
\nonumber \\
+  \frac{ \left( \left(1- \beta_\rho^2\right) s -2 m_\rho^2\right)^2 }{2 \beta_\rho  s} \log
   \left(\frac{(1-\beta_\rho ) s-2 m_\rho^2}{(1+\beta_\rho ) s-2 m_\rho^2}\right) \bigg] \nonumber \\
\eeq
for the $\rho\rho$ production inelastic amplitude with entrance channel $\pi\pi$. This amplitude is real and negative for $s>4m_\rho^2$ (note the logarithm itself is negative).

The scalar-isoscalar projection in Eq.~(\ref{rhoprod}) is shown in figure~\ref{fig:b00} with the threshold at $E_0=2m_\rho =4$ TeV ($s_0=16$ TeV$^2$). We are allowed to  vary the value of $g_s$ and $f$, but  need to  ensure $f_2^2 > 0$. The amplitude is finite at the double-$\rho$ threshold. Notice from figure~\ref{fig:b00} that  the modulus of the amplitude grows most  linearly in the perturbation theory, and we are able to have the linear term finely tuned  to avoid stringent contraint from the inelastic scattering channel.  However it  will ultimately  violate  unitarity in the large $s$ limit.  We abstain from unitarizing it at the current stage  since, at such a  energy scale, one should start  thinking  whether other resonances  should be included into the effective theory.

%%%%%%%%%%%%%%%%%%%%%%%%%%%%%%%%%%%%%%%%%%%%%%%%%%%%%%%%%%%%%%%%%%%%%%%%%%
\section{Conclusions} \label{sec:fin}
%%%%%%%%%%%%%%%%%%%%%%%%%%%%%%%%%%%%%%%%%%%%%%%%%%%%%%%%%%%%%%%%%%%%%%%%%%

In this work we have examined a simple CHM template with vector resonances assumed to be accessible at the LHC. As CHMs naturally come with a family of such new states, both charged and neutral, we have implicitly taken the pragmatic approach (in the
sense that it enabled us to perform accurate numerical studies that would otherwise not been possible) of assuming that the
CERN collider will find initially only one (degenerate) pair of such states. This could be the lowest lying one in terms of mass or else the most strongly interactive one with SM matter.

{After obtaining the scattering amplitudes among the low-energy particles, $W_L$ and $h$, their partial wave projections, and adopting the improved $K$-matrix method of unitarization, we have exposed an effective scalar resonance (equivalent to the $\sigma$ meson of low-energy hadron physics), wherein the keyword ``effective'' is meant to highlight the fact that this object, other than being a proper spin-0 state, could well appear as such yet being the scalar polarization of one or more of the additional vector resonances naturally present in CHMs but not seen at the LHC, even lighter than the detected $\rho$ states (if weakly coupled to SM objects).  In fact,
the proliferation of new gauge resonances typical of CHMs can also account for the rather broad appearance of the new $\sigma$ state, as its pole may well be just the typical mass scale of the
unseen spin-1 states and its width the linear sum of the individual ones 
(as realisable in case of multiple, nearly overlapping and typically narrower Breit-Wigner shapes with or without additional phases related to pole residues \cite{BW}), the latter taken from the coset (or even from additional
sites). This dynamics also leaves plenty of scope in a typical CHM for adequate interpretation, as it should  be recalled that the aforementioned vector resonances come accompanied in these scenarios by a variety of heavy fermions, the former decaying into the latter, with the fermionic
 masses dictating the size of the width of both the $\rho$ states and other vector bosons present in the spectrum, as emphasised in \cite{Barducci:2012kk} in one specific CHM realization, the so-called 4DCHM, which we have
used as reference benchmark herein. 
However, in this connection, a caveat should be borne in mind. 
{Since in the 4DCHM
there is no space for resonances lighter than $\rho_{L,R}$~\footnote{An exception is the gauge boson associated with an extra U(1) symmetry, necessary for the correct hypercharge assignment to the SM fermions, which however is weakly coupled to $W_L W_L$.}
which are also the most strongly coupled ones to SM matter and forces,
we can interpret our results by saying that, if we want this construct to satisfy perturbative unitarity, we must require $f$ to be  larger than  the threshold value for which $m_0>m_\rho$.
Also, we need to invoke spin-1 resonances with larger mass and width, not included in the 4DCHM, which is a two-sites truncated theory describing only the lowest lying resonances, to realize the``sigma" as a cooperative effect.}

 Whichever the underlying CHM realization though, the position in the complex $s$-plane of this resonance should depend slightly on the partial width of the $\rho$s but more strongly on the vector mass $m_\rho$ with an inverse relation, so when one becomes lighter, the other is heavier and vice versa.

Altogether, when all the relations among parameters in the CHMs at hand are taken into account, the presence of the $\rho$  at fixed $f$ improves perturbative unitarity and pushes the effective scalar pole deeper into the complex plane. It will therefore be crucial, in the case of a $\rho$ state discovery at the LHC, to closely scrutinize its properties (mass, width and couplings) in order to ascertain, through the unitarization method that we have advocated here, whether and where additional states above and beyond the SM spectrum augmented by the new visible states can be found.
We therefore conclude that, unless specific model assumptions are made on the nature of a $\rho$ state accessible at the LHC,  the unitarization procedure adopted here is a powerful method to gain substantive knowledge (mass, width, spin, etc.) of the yet unseen spectrum of the CHM at hand, either lighter or
heavier than the $\rho$ itself, in a model-independent approach. In fact, such an approach can be extended to incorporate further discoveries of both vector and scalar states that might occur at the LHC in the years to come.}

%%%%%%%%%%%%%%%%%%%%%%%%%%%%%%%%%%%%%%%%%%%%%%%%%%%%%%%%%%%%%%%%%%%%%%%%%%
\section*{Acknowledgments}
%%%%%%%%%%%%%%%%%%%%%%%%%%%%%%%%%%%%%%%%%%%%%%%%%%%%%%%%%%%%%%%%%%%%%%%%%%
FJLE thanks instructive conversations with A. Dobado, particularly
about Eq.~(\ref{secondsheet}), and support from the grant
FPA2011-27853-01 as well as computer resources, technical
expertise and assistance from the Red Espa\~nola de
Supercomputaci\'on. FJLE also thanks the Southampton High Energy
Physics (SHEP) group for hospitality at the time when this project
was conceived. SM is financed in part through the NExT Institute.
SM also acknowledges insightful discussions with R.L. Delgado.
HCai  is  supported in part by the postdoc foundation under the
Grant No. 2012M510001.

%%%%%%%%%%%%%%%%%%%%%%%%%%%%%%%%%%%%%%%%%%%%%%%%%%%%%%%%%%%%%%%%%%%%%%%%%%


\begin{thebibliography}{99}
%%%%%%%%%%%%%%%%%%%%%%%%%%%%%%%%%%%%%%%%%%%%%%%%%%%%%%%%%%%%%%%%%%%%%%%%%%

\bibitem{ATLASCMS}
G. Aad et al. (ATLAS Collaboration), Phys. Lett. B {\bf 716}, 1
(2012); S. Chatrchyan et al. (CMS Collaboration), Phys. Lett. B {\bf 716}, 30 (2012); G. Aad et al. (ATLAS Collaboration), Report No. ATLAS-CONF-2012-168; S. Chatrchyan et al.
(CMS Collaboration), Report No. CMS-HIG-12-015.


%\cite{Khachatryan:2014gha}
\bibitem{Khachatryan:2014gha}
Precise exclusion limits within specific models have been recently reported in
  V.~Khachatryan {\it et al.}  [CMS Collaboration],
  %``Search for massive resonances decaying into pairs of boosted bosons in semi-leptonic final states at $\sqrt{s}$ = 8 TeV,''
  arXiv:1405.3447 [hep-ex] and
  %%CITATION = ARXIV:1405.3447;%%
%\cite{Khachatryan:2014xja}
%\bibitem{Khachatryan:2014xja}
  V.~Khachatryan {\it et al.}  [CMS Collaboration],
  %``Search for new resonances decaying via WZ to leptons in proton-proton collisions at sqrt(s)=8 TeV,''
  arXiv:1407.3476 [hep-ex].

%\cite{Kaplan:1983fs}
\bibitem{Kaplan:1983fs}
  D.~B.~Kaplan and H.~Georgi,
  %``SU(2) x U(1) Breaking by Vacuum Misalignment,''
  Phys.\ Lett.\ B {\bf 136} (1984) 183.
  %%CITATION = PHLTA,B136,183;%%
  %393 citations counted in INSPIRE as of 19 Aug 2014

%\cite{Agashe:2004rs}
\bibitem{Agashe:2004rs}
  K.~Agashe, R.~Contino and A.~Pomarol,
  %``The Minimal composite Higgs model,''
  Nucl.\ Phys.\ B {\bf 719}, 165 (2005).
%  [hep-ph/0412089].
  %%CITATION = HEP-PH/0412089;%%


\bibitem{ET}
J.M. Cornwall, D.N. Levin and G. Tiktopoulos, Phys. Rev. D {\bf 10} (1974) 1145;
C.E. Vayonakis, Lett. Nuovo Cim. {\bf 17} (1976) 383;
B.W. Lee, C. Quigg and H. Thacker, Phys. Rev. D {\bf 16} (1977) 1519;
M.S. Chanowitz and M.K. Gaillard, Nucl. Phys. B {\bf 261} (1985) 379;
M. S. Chanowitz, M. Golden and H. Georgi, Phys. Rev. D {\bf 36} (1987) 1490;
A. Dobado J. R. Pel\'aez Nucl. Phys. B {\bf 425} (1994) 110;
Phys. Lett. B {\bf 329} (1994) 469 (Addendum, ibid, B {\bf 335} (1994) 554).



%\cite{Siringo:2001hm}
\bibitem{Siringo:2001hm}
  F.~Siringo,
  %``Light Higgs bosons from a strongly interacting Higgs sector,''
  Europhys.\ Lett.\  {\bf 57}, 820 (2002).
  %%CITATION = HEP-PH/0105018;%%

% comparison of unitarisation methods;  the new scalar boson
%\cite{Delgado:2013loa}
\bibitem{Delgado:2013loa}
  R.~L.~Delgado, A.~Dobado and F.~J.~Llanes-Estrada,
  %``Light ‘Higgs’, yet strong interactions,''
  J.\ Phys.\ G {\bf 41}, 025002 (2014).
  %%CITATION = ARXIV:1308.1629;%%

\bibitem{Bhattacharyya:2012tj}
  G.~Bhattacharyya, D.~Das and P.~B.~Pal,
  %``Modified Higgs couplings and unitarity violation,''
  Phys.\ Rev.\ D {\bf 87}, 011702 (2013).
%  [arXiv:1212.4651 [hep-ph]].
  %%CITATION = ARXIV:1212.4651;%%




\bibitem{Lahiri:2011ic}
  A.~Lahiri and D.~Mukhopadhyay,
  %``Unitarity in $WW \to WW$ elastic scattering without a Higgs boson,''
  arXiv:1107.1501 [hep-ph].
  %%CITATION = ARXIV:1107.1501;%%


\bibitem{Basso:2011na}
  L.~Basso, S.~Moretti and G.~M.~Pruna,
  %``Theoretical constraints on the couplings of non-exotic minimal $Z'$ bosons,''
  JHEP {\bf 1108}, 122 (2011).
  %%CITATION = ARXIV:1106.4762;%%


\bibitem{Contino:2011np}
  R.~Contino, D.~Marzocca, D.~Pappadopulo and R.~Rattazzi,
  %``On the effect of resonances in composite Higgs phenomenology,''
  JHEP {\bf 1110}, 081 (2011).
  %%CITATION = ARXIV:1109.1570;%%
  
%\cite{Agashe:2007ki}
\bibitem{Agashe:2007ki}
  K.~Agashe, H.~Davoudiasl, S.~Gopalakrishna, T.~Han, G.~Y.~Huang, G.~Perez, Z.~G.~Si and A.~Soni,
  %``LHC Signals for Warped Electroweak Neutral Gauge Bosons,''
  Phys.\ Rev.\ D {\bf 76} (2007) 115015
  [arXiv:0709.0007 [hep-ph]].
  %%CITATION = ARXIV:0709.0007;%%
  %160 citations counted in INSPIRE as of 05 Jan 2015
  

  
%\cite{Agashe:2008jb}
\bibitem{Agashe:2008jb}
  K.~Agashe, S.~Gopalakrishna, T.~Han, G.~Y.~Huang and A.~Soni,
  %``LHC Signals for Warped Electroweak Charged Gauge Bosons,''
  Phys.\ Rev.\ D {\bf 80} (2009) 075007
  [arXiv:0810.1497 [hep-ph]].
  %%CITATION = ARXIV:0810.1497;%%
  %86 citations counted in INSPIRE as of 05 Jan 2015  

%\cite{Barducci:2012kk}
\bibitem{Barducci:2012kk}
  D.~Barducci, A.~Belyaev, S.~De Curtis, S.~Moretti and G.~M.~Pruna,
  %``Exploring Drell-Yan signals from the 4D Composite Higgs Model at the LHC,''
  JHEP {\bf 1304} (2013) 152;
%  [arXiv:1210.2927 [hep-ph]].
  %%CITATION = ARXIV:1210.2927;%%
  %13 citations counted in INSPIRE as of 19 Aug 2014
%\cite{Barducci:2012as}
%\bibitem{Barducci:2012as}
  D.~Barducci, L.~Fedeli, S.~Moretti, S.~De Curtis and G.~M.~Pruna,
  %``Leptonic final states from di-boson production at the LHC in the 4-Dimensional Composite Higgs Model,''
  JHEP {\bf 1304}, 038 (2013).
  %[arXiv:1212.4875 [hep-ph]].
  %%CITATION = ARXIV:1212.4875;%%
  %6 citations counted in INSPIRE as of 05 Aug 2014

%\cite{DeCurtis:2011yx}
\bibitem{DeCurtis:2011yx}
  S.~De Curtis, M.~Redi and A.~Tesi,
  %``The 4D Composite Higgs,''
  JHEP {\bf 1204}, 042 (2012).
%  [arXiv:1110.1613 [hep-ph]].
  %%CITATION = ARXIV:1110.1613;%%

 %\cite{DeCurtis:2014oza}
\bibitem{DeCurtis:2014oza}
  S.~De Curtis, M.~Redi and E.~Vigiani,
  %``Non Minimal Terms in Composite Higgs Models and in QCD,''
  JHEP {\bf 1406} (2014) 071.
%  [arXiv:1403.3116 [hep-ph]].
  %%CITATION = ARXIV:1403.3116;%%
  %2 citations counted in INSPIRE as of 19 Aug 2014

%\cite{Cai:2014kxa}
\bibitem{Cai:2014kxa}
  H.~Cai,
  %``Vector-like Fermions in a Minimal Composite Higgs Model,''
  arXiv:1405.7664 [hep-ph].
  %%CITATION = ARXIV:1405.7664;%%

%\cite{Kawarabayashi:1966kd}
\bibitem{Kawarabayashi:1966kd}
  K.~Kawarabayashi and M.~Suzuki,
  %``Partially conserved g vector current and the decays of vector mesons,''
  Phys.\ Rev.\ Lett.\  {\bf 16}, 255 (1966);
  %%CITATION = PRLTA,16,255;%%
%\cite{Riazuddin:1966sw}
%\bibitem{Riazuddin:1966sw}
  Riazuddin and Fayyazuddin,
  %``Algebra of current components and decay widths of rho and K* mesons,''
  Phys.\ Rev.\  {\bf 147}, 1071 (1966);
  %%CITATION = PHRVA,147,1071;%%
  L.~v.~Dung and T.~N.~Truong,
  %``Equivalence between vector meson dominance and unitarized chiral perturbation theory,''
  hep-ph/9607378.
  %%CITATION = HEP-PH/9607378;%%

%\cite{Contino:2013gna}
\bibitem{Contino:2013gna}
  R.~Contino, C.~Grojean, D.~Pappadopulo, R.~Rattazzi and A.~Thamm,
  %``Strong Higgs Interactions at a Linear Collider,''
  JHEP {\bf 1402}, 006 (2014).
  %%CITATION = ARXIV:1309.7038;%%

\bibitem{BRST}
C. Becchi, A. Rouet and R. Stora, Phys. Lett. B {\bf 52}, 344 (1974);
C. Becchi, A. Rouet and R. Stora,  Commun. Math. Phys. {\bf 42}, 127 (1975);
 C. Becchi, A. Rouet and R. Stora,
Ann. Phys. {\bf 98}, 2 (1976);
I.V. Tyutin, arXiv:0812.0580 [hep-th].



\bibitem{Alonso:2012px}
  R.~Alonso, {\it et al.}
  %``The Effective Chiral Lagrangian for a Light Dynamical 'Higgs',''
  Phys.\ Lett.\ B{\bf 722}, 330 (2013).
%  [arXiv:1212.3305 [hep-ph]].


\bibitem{Pich:2013fba}
  A.~Pich, I.~Rosell and J.~J.~Sanz-Cillero,
  %``Strongly Coupled Models with a Higgs-like Boson,''
  arXiv:1307.1958 [hep-ph].
  %%CITATION = ARXIV:1307.1958;%%


%\cite{Degrande:2012wf}
\bibitem{Degrande:2012wf}
  C.~Degrande, N.~Greiner, W.~Kilian, O.~Mattelaer, H.~Mebane, T.~Stelzer, S.~Willenbrock and C.~Zhang,
  %``Effective Field Theory: A Modern Approach to Anomalous Couplings,''
  Ann. Phys.\  {\bf 335}, 21 (2013).
%  [arXiv:1205.4231 [hep-ph]].
  %%CITATION = ARXIV:1205.4231;%%


%\cite{Buchalla:2013rka}
\bibitem{Buchalla:2013rka}
  G.~Buchalla, O.~Cata and C.~Krause,
  %``Complete Electroweak Chiral Lagrangian with a Light Higgs at NLO,''
  arXiv:1307.5017 [hep-ph];
  %%CITATION = ARXIV:1307.5017;%%
%\cite{Buchalla:2012qq}
%\bibitem{Buchalla:2012qq}
  G.~Buchalla and O.~Cata,
  %``Effective Theory of a Dynamically Broken Electroweak Standard Model at NLO,''
  JHEP {\bf 1207}, 101 (2012).
%  [arXiv:1203.6510 [hep-ph]].
  %%CITATION = ARXIV:1203.6510;%%


\bibitem{Delgado:2013hxa}
  R.~L.~Delgado, A.~Dobado and F.~J.~Llanes-Estrada,
  %``One-loop $W_LW_L$ and $Z_LZ_L$ scattering from the electroweak Chiral Lagrangian with a light Higgs-like scalar,''
  JHEP {\bf 1402}, 121 (2014);
  %%CITATION = ARXIV:1311.5993;%%
  A.~Dobado, R.~L.~Delgado and F.~J.~Llanes-Estrada,
  %``Strongly Interacting Electroweak Symmetry Breaking Sector with a Higgs-like light scalar,''
Proc. of the II Russian-Spanish Congress "Particle and Nuclear Physics at all Scales and Cosmology" Institute of Cosmos Sciences, Saint-Petersburg, October 1-4, 2013 [arXiv:1402.0666 [hep-ph]].
  %%CITATION = ARXIV:1402.0666;%%



%GFitter precision 0.98<a^2<1.12 at 95% precision
%\cite{Baak:2012kk}
\bibitem{Baak:2012kk}
Fits of precision EW  observables lead to more stringent bounds on $a$ and thus on $K_1$ but they depend on unknown high-energy effects in virtual states, so they are not as reliable as direct LHC bounds. See
  M.~Baak {\it et al.},
  %``The Electroweak Fit of the Standard Model after the Discovery of a New Boson at the LHC,''
  Eur.\ Phys.\ J.\ C{\bf 72}, 2205 (2012).
  %%CITATION = ARXIV:1209.2716;%%

%
%\textcolor{ForestGreen}{
\bibitem{Espriu:2014jya}
  D.~Espriu and F.~Mescia,
  %``Unitarity and causality constraints in composite Higgs models,''
  Phys.\ Rev.\ D {\bf 90}, 015035 (2014).
%  [arXiv:1403.7386 [hep-ph]].
  %%CITATION = ARXIV:1403.7386;%%
%}
%\cite{Wigner:1946zz}
\bibitem{Wigner:1946zz}
  E.~P.~Wigner,
  %``Resonance Reactions and Anomalous Scattering,''
  Phys.\ Rev.\  {\bf 70}, 15 (1946);
  %%CITATION = PHRVA,70,15;%%
  E.~P.~Wigner and L.~Eisenbud,
  %``Higher Angular Momenta and Long Range Interaction in Resonance Reactions,''
  Phys.\ Rev.\  {\bf 72}, 29 (1947).
  %%CITATION = PHRVA,72,29;%%


%\cite{Oller:1999mg}
\bibitem{Oller:1999mg}
  J.~A.~Oller and E.~Oset,
  %``Two meson scattering amplitudes and their resonances from chiral symmetry and the N / D method,''
  Nucl.\ Phys.\ A {\bf 663}, 629 (2000).
%  [hep-ph/9908495].
  %%CITATION = HEP-PH/9908495;%%

%\cite{GomezNicola:2001as}
\bibitem{GomezNicola:2001as}
  A.~Gomez Nicola and J.~R.~Pelaez,
  %``Meson meson scattering within one loop chiral perturbation theory and its unitarization,''
  Phys.\ Rev.\ D {\bf 65}, 054009 (2002);
  %%CITATION = HEP-PH/0109056;%%
  A.~Dobado, M.~J.~Herrero and T.~N.~Truong,
  %``Unitarized Chiral Perturbation Theory for Elastic Pion-Pion Scattering,''
  Phys.\ Lett.\ B {\bf 235}, 134 (1990);
  %%CITATION = PHLTA,B235,134;%%
  A.~Dobado, M.~J.~Herrero and T.~N.~Truong,
  %``Study of the Strongly Interacting Higgs Sector,''
  Phys.\ Lett.\ B {\bf 235}, 129 (1990).
  %%CITATION = PHLTA,B235,129;%%

%\cite{Ananthanarayan:2000ht}
\bibitem{Ananthanarayan:2000ht}
  B.~Ananthanarayan, G.~Colangelo, J.~Gasser and H.~Leutwyler,
  %``Roy equation analysis of pi pi scattering,''
  Phys.\ Rept.\  {\bf 353}, 207 (2001).
%  [hep-ph/0005297].
  %%CITATION = HEP-PH/0005297;%%

%\cite{Kaminski:2006qe}
\bibitem{Kaminski:2006qe}
  R.~Kaminski, J.~R.~Pelaez and F.~J.~Yndurain,
  %``The Pion-pion scattering amplitude. III. Improving the analysis with forward dispersion relations and Roy equations,''
  Phys.\ Rev.\ D {\bf 77}, 054015 (2008).
%  [arXiv:0710.1150 [hep-ph]].
  %%CITATION = ARXIV:0710.1150;%%

\bibitem{Cai:2013ira} 
  H.~Cai,
  %``Higgs decay into a diphoton in the composite Higgs model,''
  Phys.\ Rev.\ D {\bf 88}, no. 3, 035018 (2013)
  [arXiv:1303.3833 [hep-ph]].

\bibitem{BW}
S. Ceci, M. Korolija and B. Zauner,
Phys. Rev. Lett. {\bf 111}, 112004 (2013).



%%%%%%%%%% PENDIENTES DE CITAR







\end{thebibliography}
\end{document}